\documentclass[aps,amsfonts,nofootinbib,onecolumn,floatfix]{revtex4}

\usepackage{hyperref}
\input epsf
\usepackage{graphics}
\usepackage{amsmath}
\usepackage{color}
\usepackage{dcolumn}

\usepackage{graphicx}

\newcommand{\beqa}{\begin{eqnarray}}
\newcommand{\eeqa}{\end{eqnarray}}

\newcommand{\bn}{\hat{\bf n}}

\newcommand{\beq}{\begin{equation}}
\newcommand{\eeq}{\end{equation}}
\newcommand{\bfl}{{\mathbf{l}}}

\newcommand{\bfL}{{\mathbf{L}}}
\newcommand{\bfLp}{{\mathbf{L^{\prime}}}}

\newcommand{\intl}[1]{\int {d^2 \bfl_{#1} \over (2\pi)^2}}

\newcommand{\vsp}{\vphantom{\Big[}\\}

\newcommand{\calb}{a}
\newcommand{\rot}{\omega}

\newcommand{\mnras}{Mon. Not. R. Astron. Soc.}

\newcommand{\araa}{Ann. Rev. Astron. Astrophys.}
\newcommand{\apss}{Astrophys. Space Sci.}

\newcommand{\cmb}{T}

\newlength{\tskip}
\newlength{\colwidth}

\def\be{\begin{equation}}
\def\ee{\end{equation}}
\def\ba{\begin{eqnarray}}
\def\ea{\end{eqnarray}}

\begin{document}

\title{Primordial B-mode Diagnostics and Self Calibrating the CMB Polarization}
\author{Amit P.S. Yadav$^{1,2}$, Meng Su$^2$, Matias Zaldarriaga$^{1,2}$}
\affiliation{$^1$ Institute for Advanced Study, School of Natural Sciences, Einstein Drive, Princeton, NJ 08540}
\affiliation{$^2$ Center for Astrophysics, Harvard University, Cambridge, MA 02138}
\begin{abstract}
Distortions in the primordial cosmic microwave background (CMB) along the line-of-sight can be modeled and described using 11 fields. These distortion fields correspond to various cosmological signals such as weak gravitational lensing of the CMB by large-scale structure, screening from patchy reionization, rotation of the plane of polarization due to magnetic fields or parity violating physics. Various instrumental systematics such as gain fluctuations, pixel rotation, differential gain, pointing, differential ellipticity are also described by the same distortion model. All these distortions produce B-mode that contaminate the primordial tensor B-modes signal. In this paper we show that apart from generating B-modes, each distortion uniquely couples different modes $(\bfl_1\ne \bfl_2)$ of the CMB anisotropies, generating $\langle EB \rangle$ correlations which for the primordial CMB are zero. We describe and implement unbiased minimum variance quadratic estimators which using the off diagonal correlations in the CMB can extract the map of  distortions.  We perform Monte-Carlo simulations to characterize the estimators and illustrate the level of distortions that can be detected with current and future experiments. The estimators can be used to look for cosmological signals, or to check for any residual systematics in the data. As a specific example of primordial tensor B-mode diagnostics we compare the level of minimum detectable distortions using our method with maximum allowed distortion level for the B-modes detection. We show that for any experiment, the distortions will be detected at high significance using correlations before they would show up as spurious B-modes in the power spectrum. In another words if B-modes were to be detected, the presence of any non-primordial contribution could be checked and removed using the off-diagonal correlations in the CMB.  
\end{abstract}
\maketitle
\section{Introduction}
Inflation is perhaps the most promising paradigm of the early universe. Inflation resolves various problems with the standard Big-Bang scenario (see~\cite{baumann_cmbpol,Baumann:2009ds} for reviews) and provides seed scalar perturbations for structure formation. Besides the scalar perturbations, inflation also predicts tensor perturbations or gravitational waves. Detection of these tensor perturbations is considered as a ``smoking gun'' for inflationary scenarios. Primordial scalar perturbations create only E-modes of the CMB\footnote{To first order in perturbations, primordial scalar perturbations do not generate B-modes of CMB. However at second (and higher) order in perturbations, scalar perturbations do produce B-modes~\cite{Bartolo:2006fj,Baumann:2007zm}. The B-modes generated from higher order perturbations are expected to be smaller than the tensor B-mode levels that the upcoming and future experiments (like CMBPol) are sensitive too.}, while primordial tensor perturbations generate both parity even E-modes and parity odd B-modes polarization~\cite{SeljakZaldarriaga97,1997PhRvD..55.7368K,1997PhRvL..78.2058K}. The detection of primordial tensor B-modes~\footnote{Throughout the paper the term primordial B-modes refers to the B-modes generated during inflation via tensor perturbations.} in the CMB would confirm the existence of tensor perturbations in the early universe. Various observational efforts are underway to detect such B-mode signal of the CMB~\cite{ebex,cmb_polarization_experiments}. The magnitude of tensor generated B-mode signal is directly related to the Hubble parameter $H$ during inflation, and thus a detection would establish the energy scale at which inflation happened. The amplitude of primordial B-modes can be characterized as a ratio of tensor-to-scalar perturbation amplitudes, $r$. The limit from WMAP 5-year data is $r < 0.22$~\cite{Komatsu2009}.

Apart from foreground subtraction challenges, and the challenge of reaching the instrumental sensitivity to detect primordial B-modes, there are also many sources of non-primordial B-modes which contaminate the inflationary signal. Some example physical processes that generate B-modes are, weak lensing of the CMB by large-scale structure~\cite{1996ApJ...463....1S,Zaldarriaga:1998ar}, patchy re-ionization~\cite{2000ApJ...529...12H,DvorkinHuSmith}, rotation of the CMB polarization~\cite{Kamionkowski2009,Yadav_etal2009}. Instrumental systematics also generate spurious B-modes~\cite{HHZ,Shimon2007}. What if B-modes are detected? How would we convince ourselves that the B-modes are primordial? In this paper we show that any line-of-sight ``distortion'' in the primordial CMB that generates non-primordial B-modes also leaves a distinct signature in the otherwise zero $\langle EB\rangle$ and $\langle TB \rangle$ correlations~\cite{SeljakZaldarriaga97,1997PhRvD..55.7368K,1997PhRvL..78.2058K,2009PhRvD..79j3002M}. We show that the amplitude of $\langle EB \rangle$ (and TB) correlations is proportional to the distorting field while the B-mode power spectrum is proportional to the power spectrum of the distorting field. We construct an unbiased minimum variance quadratic estimator for the distortion field ${\cal D}$ using the quadratic combination of observed $E,B$ and $T$ CMB maps. We find that if an experiment sees non-primordial B-modes generated by distortion ${\cal D}$, our estimator would detect it at higher significance.

The organization of the paper is as follows: In Sec.~\ref{sec:distortions} we introduce a model for describing the distortions in the CMB, and discuss the effect of distortions on the CMB fields and their correlations. In Sec.~\ref{sec:estimator} we construct and characterize a minimum variance estimator which can reconstruct the distortion fields. In Sec.~\ref{sec:correspondence} we show the correspondence between our 11 distortion fields with the known physical and instrumental effects on the CMB. In Sec.~\ref{sec:results} we discuss numerical results and discuss the use of our formalism, with an emphasis on primordial B-mode diagnostics. Finally in Sec.~\ref{sec:summary} we summarize our findings.

\section{Distortions in the primordial CMB maps}
\label{sec:distortions}
We parametrize the distortions in the CMB polarization map following Ref.~\cite{HHZ}. Changes in the Stokes parameter of the CMB due to distortions along the line-of-sight can be written as
\begin{eqnarray}
\delta [Q \pm i U](\bn) &=&
            [\calb \pm i 2 \rot](\bn)  [\tilde Q \pm i \tilde U](\bn) + [f_1 \pm i f_2](\bn)   [\tilde Q \mp i \tilde U](\bn) + [\gamma_1 \pm i \gamma_2](\bn) \tilde \cmb(\bn) \nonumber \\
 &&+\sigma {\bf p}(\bn) \cdot \nabla [\tilde Q \pm i \tilde U](\bn;\sigma) + \sigma [d_1 \pm i d_2](\bn) [\partial_1 \pm i\partial_2] \tilde T(\bn;\sigma)
 + \sigma^2 q(\bn) [\partial_1 \pm i \partial_2]^2 \tilde \cmb(\bn;\sigma)+\ldots \,.
\label{eqn:model_distortions}
\end{eqnarray}
The first line  captures the distortions in a  single perfectly known direction $\bn$. The distortions in second line capture mixing of the polarization fields in a local region of length scale $\sigma$ around $\bn$. We Taylor expand the CMB fields $Q, U,$ and $ T$ around the point $\bn$ and consider the leading order terms. Here $\tilde Q, \tilde U,$ and $\tilde T$ stands for primordial (un-distorted) CMB fields. Since $(Q \pm iU)(\bn)$ is spin $\pm 2$ field, $a(\bn)$ is a scalar field that describes modulation in the amplitude of the fields in a given direction $\bn$; $\rot(\bn)$ is also a scalar field that describes the rotation of the plane of polarization, $(f_1\pm if_2)$ are spin $\pm 4$ fields that describe the coupling between two spin states (spin-flip),
 and $(\gamma_1\pm i \gamma_2)(\bn)$ are spin $\pm2$ fields that describe leakage from the temperature to polarization (monopole leakage hereon). Distortions in the second line of Eqn.~(\ref{eqn:model_distortions}), $(p_1\pm p_2), (d_1 \pm d_2)$, and $q$ are measured in the units of the length scale $\sigma$.
The field $(p_1\pm ip_2)(\bn)$ is a spin $\pm 1$ field and describes the change in the photon direction; we will refer to it as a deflection field. Finally $(d_1\pm d_2)(\bn)$ and $q(\bn)$ describe leakage from temperature to polarization, $(d_1\pm d_2)(\bn)$ is spin $\pm1$ field and we will refer to it as dipole leakage; $q(\bn)$ is a scalar field that we will call quadrupole leakage.

These 11 distortion fields correspond to various cosmological distortions such as weak gravitational lensing of the CMB, screening effects from patchy reionization, rotation of the plane of polarization due to magnetic fields or parity violating physics and various instrumental systematics such as gain fluctuations, pixel rotation, differential gain, pointing, differential ellipticity are also described by our distortion model. In Sec.~\ref{sec:correspondence} we will discuss the correspondence between the distortion fields and instrumental and cosmological signals in more detail.

We will work in the flat-sky limit where scalar fields such as the CMB temperature $T$ and a complex field $({\cal S}_1 \pm i {\cal S}_2)(\bn)$ of spin $\pm s$ can be expanded in the Fourier basis as
\begin{eqnarray}
T(\bfl) &=& \int d \bn \, T(\bn) e^{-i \bfl \cdot \bn}\,, \nonumber \\
\left[{\cal S}_1 \pm i {\cal S}_2 \right] (\bfl) &=& (\pm 1)^s \int  d \bn\, [{\cal S}_1(\bn)\pm i {\cal S}_2(\bn)] e^{\mp si\varphi_{\bf l}} e^{-i \bfl \cdot \bn},
\label{EBFields}
\end{eqnarray}
where $\varphi_{\bfl}=\cos^{-1}({\bn} \cdot \hat \bfl)$. The complex field $(Q\pm iU)(\bn)$ is a spin $\pm 2$ field, whose Fourier harmonics are referred as $(E\pm iB)(\bfl)$.  
We want to calculate the change in the CMB field due to the distortion field ${\cal D}(\bn)$. 

 We Taylor expand the observed CMB fields assuming that the distortions are small, keeping terms only to leading order in the distortion field ${\cal D}$. The changes in the $E$ and $B$ fields due to the modulation $a$, rotation $\rot$, spin-flip $(f_1,f_2)$
and deflection $(p_1,p_2)$ take the following form 
\begin{eqnarray}
B(\bfL) &=& \int {d^2 \bfl_1 \over (2\pi)^2} {\cal D}(\bfl_1) \tilde E(\bfl_2) W^B_{\cal D}(\bfl_1,\bfl_2)\,, \nonumber \\
E(\bfL) &=& \tilde E(\bfL) + \int {d^2 \bfl_1 \over (2\pi)^2} {\cal D}(\bfl_1) \tilde E(\bfl_2) W^E_{\cal D}(\bfl_1,\bfl_2)\,,
\label{eqn:enb1}
\end{eqnarray}
where $\bfl_2 = \bfL - \bfl_1$ and $W^B_{\cal D}$, $W^{E}_{\cal D}$ are given in Table~\ref{tab:filters}.  We have assumed zero primordial B-modes for our fiducial model. Similarly the changes due to temperature leakage, $(\gamma_1,\gamma_2),q$, and $(d_1,d_2)$ can be described by
\begin{eqnarray}
B(\bfL) &=& \int {d^2 \bfl_1 \over (2\pi)^2} {\cal D}(\bfl_1) \tilde T(\bfl_2) W^B_{\cal D}(\bfl_1,\bfl_2)\,, \nonumber \\
E(\bfL) &=& \tilde E(\bfL) +  \int {d^2 \bfl_1 \over (2\pi)^2} {\cal D}(\bfl_1) \tilde T(\bfl_2) W^E_{\cal D}(\bfl_1,\bfl_2)\,.
\label{eqn:enb2}
\end{eqnarray}
It is clear from Eqs.~(\ref{eqn:enb1}) and~(\ref{eqn:enb2}) that the distortion field of wavevector $\bfL$ mixes the polarization
modes of wavevectors $\bfl_1$ and $\bfl_2=\bfL -\bfl_1$. 

The power spectra of the different modes of the primordial CMB field are defined as
\begin{eqnarray}
\langle \tilde X(\bfl_1) \tilde X'(\bfl_2)\rangle &=&(2\pi)^2\delta(\bfl_1+\bfl_2)\tilde C^{XX'}_{\ell_1}\,,
\end{eqnarray}
where $X,X'=T,E,B$ and
$ \langle \tilde E(\bfl_1) \tilde B(\bfl_2)\rangle = 0, \langle \tilde T(\bfl_1) \tilde B(\bfl_2)\rangle = 0\,. $
The off diagonal terms of the power spectrum $(\bfl_1 \ne \bfl_2)$ are zero because of statistical isotropy and the $\langle \tilde E(\bfl_1)\tilde B(\bfl_2) \rangle$ and $\langle \tilde T(\bfl_1)\tilde B(\bfl_2)\rangle$ correlations are zero because $\tilde E(\bfl)$ is parity even, $\tilde B(\bfl)$ is parity odd, and the physical processes responsible for the generation of the CMB anisotropies are parity conserving. The observed CMB power spectra in the presence of distortion and lensing effects can be written as

\begin{table*}[t]
\caption{Filters, $f^{\cal D}_{XX'}(\bfl_1,\bfl_2)$}
\begin{center}
\begin{tabular}{c|cccc}
\hline \hline
     ${\cal D}$          & $f^{\cal D}_{EB}({\bfl_1,\bfl_2})$ &$f^{\cal D}_{TB}({\bfl_1,\bfl_2})$ &$W^B_{{\cal D}}({\bfl_1,\bfl_2})$&$W^E_{{\cal D}}({\bfl_1,\bfl_2})$   \vsp \hline
$a$      & $\tilde C_{l_1}^{E E}\sin 2(\varphi_{\bfl_1}-\varphi_{\bfl_2})$ & $\tilde C_{l_1}^{T E}\sin 2(\varphi_{\bfl_1}-\varphi_{\bfl_2})$ & $\sin[2 (\varphi_{\bfl_2} - \varphi_\bfL)]$ & $\cos[2 (\varphi_{\bfl_2} - \varphi_\bfL)]$ \vsp 
$\rot$    & $2\tilde C^{EE}_{l_1}\cos2(\varphi_{\bfl_1} -\varphi_{\bfl_2})$ & $2\tilde C^{TE}_{l_1}\cos2(\varphi_{\bfl_1} -\varphi_{\bfl_2})$ & $2\cos[2 (\varphi_{\bfl_2} - \varphi_{\bfL})]$  & $ -2 \sin[2 (\varphi_{\bfl_2} - \varphi_{\bfL})]$\vsp
$\gamma_1$      & $\tilde C^{TE}_{l_1}\sin2(\varphi_{\bfL }-\varphi_{\bfl_2})$ & $\tilde C^{TT}_{l_1}\sin2(\varphi_{\bfL }-\varphi_{\bfl_2})$& $\sin[2 (\varphi_{\bfl_1}- \varphi_\bfL)]$\,, & $\cos[2 (\varphi_{\bfl_1}- \varphi_\bfL)]$\vsp
$\gamma_2$  & $\tilde C^{TE}_{l_1}\cos 2(\varphi_{\bfL }-\varphi_{\bfl_2})$ & $\tilde C^{TT}_{l_1}\cos 2(\varphi_{\bfL }-\varphi_{\bfl_2})$& $ \cos[2 (\varphi_{\bfl_1}- \varphi_\bfL)]$\,, & $  -\sin[2 (\varphi_{\bfl_1}- \varphi_\bfL)]$\vsp
$f_1$ & $\tilde C^{EE}_{l_1}\sin2(2\varphi_{\bfL}-\varphi_{\bfl_1}-\varphi_{\bfl_2})$ & $\tilde C^{TE}_{l_1}\sin2(2\varphi_{\bfL}-\varphi_{\bfl_1}-\varphi_{\bfl_2})$&   $\sin[2 (2 \varphi_{\bfl_1} - \varphi_{\bfl_2} -\varphi_{\bfL}) ]$  & $ \cos[2 (2 \varphi_{\bfl_1} - \varphi_{\bfl_2} -\varphi_{\bfL}) ]$ \vsp
$f_2$ & $\tilde C^{EE}_{l_1}\cos2(2\varphi_{\bfL}-\varphi_{\bfl_1}-\varphi_{\bfl_2})$ &  $\tilde C^{TE}_{l_1}\cos2(2\varphi_{\bfL}-\varphi_{\bfl_1}-\varphi_{\bfl_2})$&$\cos2(2\varphi_{\bfl_1}-\varphi_{\bfl_2}-\varphi_{\bfL})$ &$-\sin2(2\varphi_{\bfl_1}-\varphi_{\bfl_2}-\varphi_{\bfL})$ \vsp
$d_1$ & $\tilde C^{TE}_{l_1} (l_1 \sigma) \cos(\varphi_{\bfL}+\varphi_{\bfl_1}-2\varphi_{\bfl_2})$ & $\tilde C^{TT}_{l_1} (l_1 \sigma) \cos(\varphi_{\bfL}+\varphi_{\bfl_1}-2\varphi_{\bfl_2})$& $     - (\bfl_2 \sigma) \cos[ \varphi_{\bfl_1} + \varphi_{\bfl_2} - 2 \varphi_l]$ & $ - (\bfl_2 \sigma) \sin[ \varphi_{\bfl_1} + \varphi_{\bfl_2} - 2 \varphi_L]$\vsp
$d_2$ &  $-\tilde C^{TE}_{l_1} (l_1 \sigma) \sin(\varphi_{\bfL}+\varphi_{\bfl_1}-2\varphi_{\bfl_2})$ & $-\tilde C^{TT}_{l_1} (l_1 \sigma) \sin(\varphi_{\bfL}+\varphi_{\bfl_1}-2\varphi_{\bfl_2})$&$ (l_2 \sigma) \sin[ \varphi_{\bfl_1} + \varphi_{\bfl_2} - 2 \varphi_{\bfL}]$ & $  (l_2 \sigma) \cos[ \varphi_{\bfl_1} + \varphi_{\bfl_2} - 2 \varphi_{\bfL}]$ \vsp
 $q$ & $-\tilde C^{TE}_{l_1} (l_1 \sigma)^2 \sin2(\varphi_{\bfl_1}-\varphi_{\bfl_2})$ &$-\tilde C^{TT}_{l_1} (l_1 \sigma)^2 \sin2(\varphi_{\bfl_1}-\varphi_{\bfl_2})$ &$  - (l_2 \sigma)^2 \sin[ 2 (\varphi_{\bfl_2} - \varphi_{\bfL})]$ & $ - (l_2 \sigma)^2 \cos[ 2 (\varphi_{\bfl_2} - \varphi_{\bfL})]$\vsp
$p_1$ & $-\tilde C^{EE}_{l_1} \sigma (\bfl_1 \times \hat \bfL) \sin2(\varphi_{\bfl_1}-\varphi_{\bfl_2})$ &  $-\tilde C^{TT}_{l_1} \sigma (\bfl_1 \times \hat \bfL) \sin2(\varphi_{\bfl_1}-\varphi_{\bfl_2})$ &   $\sigma (\bfl_2 \times \hat \bfl_1)\cdot \hat{\bf z} \sin[ 2(\varphi_{\bfl_2}- \varphi_{\bfL}) ]$ &   $\sigma (\bfl_2 \cdot \hat \bfl_1)\sin[ 2(\varphi_{\bfl_2}- \varphi_\bfL) ]$\vsp
$p_2$ & $-\tilde C^{EE}_{l_1} \sigma(\bfl_1 \cdot \hat \bfL) \sin2(\varphi_{\bfl_1}-\varphi_{\bfl_2})$ &  $-\tilde C^{TT}_{l_1} \sigma(\bfl_1 \cdot \hat \bfL) \sin2(\varphi_{\bfl_1}-\varphi_{\bfl_2})$ &$\sigma (\bfl_2  \cdot \hat \bfl_1) \sin[ 2(\varphi_{\bfl_2} - \varphi_\bfL)]$ & $ \sigma (\bfl_2  \times \hat \bfl_1) \cdot \hat{\bf z}\sin[ 2(\varphi_{\bfl_2} - \varphi_\bfL)]$ \vsp
\hline\hline

\end{tabular}
\end{center}
\label{tab:filters}
\end{table*}

\begin{eqnarray}
\langle X(\bfl_1) X'(\bfl_2)\rangle &=&(2\pi)^2\delta(\bfl_1+\bfl_2) \big[ C^{XX'}_{\ell_1} + {\cal N}^{XX'}_{\ell_1}\big] \,, 
\end{eqnarray}
with
\begin{eqnarray}
{\cal N}^{TT}_{\ell} &=&\Delta^2_T \, e^{\ell (\ell+1)\sigma^2/8\ln2}\,, \nonumber \\
{\cal N}^{EE}_{\ell}&=&{\cal N}^{BB}_{\ell} =\Delta^2_P \,e^{\ell (\ell+1)\sigma^2/8\ln2}\,, 
\end{eqnarray}
where $\sigma$ is the FWHM of the beam and $\Delta_{T}$ and $\Delta_{P}$ are the temperature and polarization detector noise, respectively. In our numerical calculations we assumed fully polarized detectors, for which $\Delta_P=\sqrt{2}\Delta_T$. 
Using Eqn.~(\ref{eqn:enb1}) and~(\ref{eqn:enb2}) we can calculate $\langle X(\bfl_1) X'(\bfl_2)\rangle$ correlations. Taking the ensemble average over the CMB fields for a fixed distortion field ${\cal D}(\bn)$, one gets
\begin{eqnarray}
\langle X(\bfl_1) X'(\bfl_2) \rangle_{\rm CMB} = f^{{\cal D}}_{XX'}(\bfl_1,\bfl_2) {\cal D}(\bfl_1+\bfl_2)\,,
\label{eqn:cxy}
\end{eqnarray} 
for $XX'=TT,TE,TB,EB$  and 
\begin{equation}
\langle B(\bfl_1) B(\bfl_2) \rangle_{\rm CMB, \cal{D}} = (2\pi)^2 \delta(\bfl_1+\bfl_2) \Bigg\{
\begin{array}{cc}
\displaystyle \int  \frac{d^2\bfl'}{(2\pi)^2} C^{{\cal D}{\cal D}}_{l'} \tilde C^{EE}_{l''} W^{B}_{\cal D}(\bfl',\bfl'') W^{B}_{\cal D}(\bfl',\bfl'') & \text{ ${\cal D}={a,\rot,(f_1,f_2),(p_1,p_2)}$}\,,\vspace{.2cm}\\
 \displaystyle \int \frac{d^2\bfl'}{(2\pi)^2}C^{{\cal D}{\cal D}}_{l'}  \tilde C^{TT}_{l''} W^{B}_{\cal D}(\bfl',\bfl'')W^{B}_{\cal D}(\bfl',\bfl'')& \hspace{.3cm}\text{${\cal D}=$ temperature leakage}\,, \\
\end{array}
\label{eqn:cbb}
\end{equation}
where $\bfl''=\bfl_1-\bfl'$ and we have assumed zero primordial B-modes for our fiducial model. The filters $f^{{\cal D}}_{XX'}(\bfl_1,\bfl_2)$ are given in Table~\ref{tab:filters}. Note that $\langle ... \rangle_{\rm CMB}$ means an average over CMB realizations. The leading order effect of distortions on the $\langle E(\bfl_1)B(\bfl_2)\rangle$ and $\langle T(\bfl_1)B(\bfl_2)\rangle$ correlations is to introduce nonzero off-diagonal terms, while for $\langle B(\bfl_1)B(\bfl_1)\rangle$ the leading order effect is to generate diagonal terms. 

\section{Minimum Variance Estimators for Distortion Fields}
\label{sec:estimator}
Our goal is to use Eqn.~(\ref{eqn:cxy}) to construct a suitable
estimator for the Fourier components ${\cal D}(\bfL)$ of the distortion
field in terms of the observed fields $T (\bfl), E (\bfl),$ and $B(\bfl)$. Following~\cite{HuOkamoto}, we can
define a minimum variance quadratic estimator $\hat{{\cal D}}_{XX'}(\bfL)$
for ${\cal D}(\bfL)$ by weighting quadratic combinations of different CMB fields $X$ by window functions $F^{{\cal D}}_{XX'}(\bfl_1,\bfl_2)$,
\begin{eqnarray}
\hat {\cal D}_{XX'}({\bfL})&=& A^{\cal D}_{XX'}(L) \intl{1}
X(\bfl_1) X'(\bfl_2) F^{{\cal D}}_{XX'}(\bfl_1,\bfl_2)\,,
\label{eqn:estimator}
\end{eqnarray}
where $\bfL=\bfl_2 + \bfl_1$, and the normalization
\begin{eqnarray}
A^{\cal D}_{XX'}(L) = \Bigg[ \intl{1} f^{\cal D}_{XX'}(\bfl_1,\bfl_2)
F^{\cal D}_{XX'}(\bfl_1,\bfl_2) \Bigg]^{-1} \,,
\label{eq:noise}
\end{eqnarray}
is chosen to make the estimator unbiased, i.e. $\langle \hat {\cal D}(\bfL)\rangle_{\rm CMB}={\cal D}(\bfL)$. 
The weights $F^{{\cal D}}_{XX'}$ can be optimized by minimizing the variance subjected to the
normalization constraint
\begin{eqnarray}
\frac{\partial}{\partial F^{\cal D}_{XX'}(\bfl_1,\bfl_2)}\Bigg\langle \Big|\hat {\cal D}_{XX'}- {\cal D}\Big|^2\Bigg\rangle_{\text{CMB, distortion}}=0\,.
\end{eqnarray}
The angle bracket stands for average over both the CMB and the distortion fields. For $XX' = TT, EE$ and $BB$, the minimization yields

\begin{equation}
F^{\cal D}_{XX}(\bfl_1,\bfl_2)= {f^{\cal D}_{XX}(\bfl_1,\bfl_2) \over
	 2 C_{\ell_1}^{XX} C_{\ell_2}^{XX}}\,.
\end{equation}
For $XX' = TB$ and $EB$, for which $\tilde C_{\ell}^{XX'}=0$, we get 
\begin{equation}
F^{\cal D}_{XX'}(\bfl_1,\bfl_2) = {f^{\cal D}_{XX'}(\bfl_1,\bfl_2) \over 
	C_{\ell_1}^{XX} C_{\ell_2}^{X'X'}},
\end{equation}
and for $XX'=TE$ we find
\begin{equation}
F^{\cal D}_{TE}(\bfl_1,\bfl_2) = {f^{\cal D}_{TE}(\bfl_1,\bfl_2) \over 
	C_{\ell_1}^{TT} C_{\ell_2}^{EE}+	C_{\ell_1}^{TE} C_{\ell_2}^{TE} },
\end{equation}
where $C_{\ell_2}^{XX}$ and $C_{\ell_2}^{X'X'}$ are the observed power spectra
including the effects of both the signal and the noise.

The variance of the estimator is
\begin{eqnarray}
\langle \hat {\cal D}_{XX'}(\bfL) \hat {\cal D}_{YY'}(\bfL^{\prime})\rangle_{\text{CMB, distortion}}= (2\pi)^2 \delta(\bfL + \bfLp) \Big\{ C^{{\cal D}{\cal D}}_L +N^{\cal D}_{XX',YY'}(L) + N^{(1){\cal D},}_{XX',YY'}(L) + N^{(2),{\cal D}}_{XX',YY'}(L)+...\Big\}\,, \nonumber \\
\label{eqn:variance}
\end{eqnarray}
where 
\begin{eqnarray}
N^{\cal D}_{XX',YY'}(L)=(2\pi)^2 A^{\cal D}_{XX'}(L)A^{\cal D}_{YY'}(L) \int \frac{d^2 {\bfl_1}}{(2\pi)^2} F^{\cal D}_{XX'}(\bfl_1,\bfl_2) \Big[F^{\cal D}_{YY'}(\bfl_1,\bfl_2)C^{XY}_{\ell_1}C^{X'Y'}_{\ell_2} + F^{\cal D}_{YY'}(\bfl_2,\bfl_1)C^{XY'}_{\ell_1}C^{X'Y}_{\ell_2}\Big]\,,
\label{eqn:GN}
\end{eqnarray}
with $\bfL=\bfl_1+\bfl_2$. For the minimum variance estimator, the Gaussian noise
$N^{\cal D}_{XX^\prime, XX'}(L)\equiv N^{{\cal D}}_{XX'}(L)= A^{{\cal D}}_{XX'}(L)$ provides the dominant contribution to
the variance. The higher order noise terms $N^{(n), {\cal D}}_{XX',YY'}(L)$ come from the connected part of the trispectrum and are proportional to $\big( C^{{\cal D}{\cal D}}(L) \big)^n$. To a good approximation these non-Gaussian noise terms are much smaller than the Gaussian noise $N^{\cal D}_{XX'}(L)$. In Fig.~(\ref{fig:N-type2}) we show the Gaussian noise $N^{\cal D}_{EB}(L)$ for all the 11 distortion fields, we discuss these plots in detail in Sec.~\ref{sec:results}.
Although for the experiments like CMBPol, $EB$ estimator is the most sensitive, for experiments such as PLANCK which has lower sensitivity to polarization, one can combine all the estimators $(TT,TE,EE,EB,TB)$ to construct a minimum variance estimator.

{\it Signal-to-Noise}: To calculate the signal-to-noise we need to assume a form for the power spectrum of the distortion field $C^{{\cal D}{\cal D}}(L)$. Motivated by the discussion in ~\cite{HHZ,SYZ09}, we assume that the distortion
fields, as defined in Eqn.~(\ref{eqn:model_distortions}),
are statistically isotropic and Gaussian. Thus their statistical
properties are fully described by their power spectra,
\begin{equation}
\left< {\cal D} (\bfl) {\cal D}(\bfl') \right> = (2\pi)^2 \delta(\bfl+\bfl') C_l^{{\cal D}{\cal D}}\,.
\end{equation}
 For illustration we assume power spectra of the form
\begin{equation}
C_l^{{\cal D}{\cal D}} = A^2_{\cal D} \exp(-l(l+1)\alpha_{{\cal D}}^2/8\ln2), \label{eqn:css}
\end{equation}
i.e. white noise above a certain coherence scale $\alpha_{{\cal D}}$. The parameter $A_{\cal D}$ characterizes the $\it rms$ of the distortion field ${\cal D}$. The signal-to-noise for a given distortion field can be calculated as
\begin{equation}
\frac{\text{Signal}}{\text{Noise}}=\Bigg[ \frac{f_{sky}}{2}\sum_{\ell} (2\ell+1)\Bigg(\frac{C^{{\cal D}{\cal D}}_\ell}{N^{{\cal D}{\cal D}}_\ell}\Bigg)^2 \Bigg]^{1/2},
\label{eqn:s2n}
\end{equation}
where $C^{{\cal D}{\cal D}}_\ell$ is the distortion power spectrum for a distortion ${\cal D}$ given by Eqn.~(\ref{eqn:css}), $N^{{\cal D}{\cal D}}(\ell)$ is the Gaussian noise for the estimator of the distortion ${\cal D}$ (given by Eqn.~(\ref{eqn:GN})). In Fig.~\ref{fig:s2n} we show total signal-to-noise for all the 11 distortions as a function of $\ell_{max}$. In Sec.~\ref{sec:results} we discuss the plot in detail.

\section{Correspondence Between the Distortion fields and Cosmological and Instrumental Effects}
\label{sec:correspondence}
\subsection{Cosmological Signals}
The distortions defined by Eqn.~(\ref{eqn:model_distortions}) are general distortions along the line-of-sight of the CMB. Here we consider known physical processes which generate such distortions in the CMB.

{\it Cosmological rotation:} The plane of linear polarization of CMB field can be 
rotated due to interactions which introduce a different 
dispersion relation for the left and right circularly 
polarized modes, during propagation from the surface of last-scattering to the Earth.  
Such interactions can come from three
main sources: (a) interactions with dust foregrounds, (b) Faraday rotation 
due to interactions with background magnetic fields, and (c) interactions with 
pseudoscalar fields~\cite{Carroll98}. The interactions with foregrounds leads to a  
frequency-dependent effect. The same is true of Faraday rotation, where the frequency dependence
is $\sim \nu^{-2}$~\citep{Kosowsky_Loeb1996,2005PhRvD..71d3006K, 2004ApJ...616....1C,2004PhRvD..70f3003S}, while the interactions with pseudo-scalar fields are frequency-independent. These rotations change the CMB Stokes parameters as
\begin{eqnarray}
(Q\pm iU)(\bn) &=&e^{-i\alpha(\bn)} (\tilde Q \pm i\tilde U)(\bn) \quad \Rightarrow \quad \delta(Q\pm iU)(\bn) =-i\alpha(\bn)(\tilde Q \pm i\tilde U)(\bn) \,,
\label{eqn:correspondence_rotation}
\end{eqnarray}
where $\alpha(\bn)$ is the integrated rotation of the plane of polarization along the line-of-sight $\bn$. Comparing Eqn.~(\ref{eqn:correspondence_rotation}) with Eqn.~(\ref{eqn:model_distortions}) it is clear that the cosmological rotation is captured in our rotation distortion field $\rot({\bn})$. Currently there are no direct constraints on spatially varying rotation, however indirect constraints from the upper limit of B-modes exist~\cite{prs08}. The constant rotation has been constrained to be less than few degrees~\cite{quad_parity,Komatsu2009}. From the QUAD data~\cite{quad_parity} the constant rotation has been constrained to
$0.55\pm0.82$ (random)$\pm0.5$ (systematic) degree. Planck can reach the
sensivity of $0.1$deg, while futuristic experiment like CMBPol will be
able to detect rotation angle as small as $0.005$ degree~\cite{Yadav_etal2009,Gluscevic:2009mm}.

{\it Patchy reionization:} Inhomogeneous reionization introduces several effects in the CMB map. First, ionized materials scatter the local
CMB temperature quadrupole, thus generating extra polarization on {\it
  large-scales}~\cite{GruzinovHu,Hu2000}. Second, extra temperature and polarization
anisotropies are generated from peculiar motion of ionized regions
during the entire reionization process, introducing a kinetic 
Sunyaev-Zeldovich signal~\cite{SZ1970,SZ1980,McQuinn2005,Santos2003,Iliev2007,Zhang2004}. Finally, the primordial
anisotropies are suppressed by a factor of $e^{-\tau(\bn)}$, where
$\tau(\bn)$ is the line-of-sight dependent optical depth to
recombination. Scattering which mixes photons from different directions, is not a line-of-sight distortion and is not modeled in Eqn.~(\ref{eqn:model_distortions}), while the {\it screening}
effect from patchy reionization is a line-of-sight effect and is indeed captured. The observed polarization Stokes parameters due to the screening effect are 
\begin{eqnarray}
(Q\pm iU)(\bn) &=&e^{-\tau(\bn)} (\tilde Q \pm i\tilde U)(\bn) \Rightarrow \delta(Q\pm iU)(\bn) =-\tau(\bn)(\tilde Q \pm i\tilde U)(\bn) \,.
 \end{eqnarray} 
Comparing the above equation with Eqn.~(\ref{eqn:model_distortions}) shows that the screening effect from patchy reionization is captured by our modulation parameter $a({\bn})$. For typical patchy reionization models, $\delta \tau(\bn)$ is of the
order of $10^{-3}$ and hence it will be challenging to reconstruct it. The corresponding
B-modes are about an order of magnitude smaller than the lensing induced B-modes~\cite{2000ApJ...529...12H,2009PhRvD..79d3003D}.

{\it Lensing:} The path of the CMB photons get deflected by the intervening matter between us and the last-scattering surface. The remapping of CMB photons due to this so called weak gravitational lensing of the CMB can be written as~\cite{Zaldarriaga:1998ar,Hirata:2003ka,Cooray:2005hm}
\begin{eqnarray}
(Q\pm iU)(\bn) &=& (\tilde Q \pm i\tilde U)(\bn+\nabla \Phi +\nabla \times \Omega)\,,
\end{eqnarray}
which after Taylor expansion can be written as  
\begin{eqnarray} 
\delta(Q\pm iU)(\bn) =\nabla\Phi \cdot \nabla(\tilde Q \pm i\tilde U)(\bn) + \nabla \times \Omega \cdot \nabla(\tilde Q \pm i\tilde U)(\bn)+\ldots \,.
\label{eqn:lensing}
\end{eqnarray}
Here $\Phi$ is a weighted projection of the gravitational potential $\Psi$ along the line-of-sight and is written as
\begin{eqnarray} 
\Phi=-2\int dD \frac{(D_\star-D)}{DD_\star}\Psi(D\bn,D)\,,
\end{eqnarray}
where $D_\star$ is the comoving distance to the last-scattering-surface. The gradient part of lensing, $\nabla \Phi$, is generated by the linear order density perturbations, while the curl part,  $\nabla \times \Omega$, comes from the second order density perturbations. The curl part is much smaller than the gradient part and is usually ignored. The form of Eqn.~(\ref{eqn:lensing}) when compared with Eqn. (\ref{eqn:model_distortions}) shows that the gradient type of lensing is captured in our deflection field $p_2(\bn)$, while the curl part of lensing is same as our deflection distortion $p_1({\bn})$.


{\it Other cosmological distortions}:
Although there are several other CMB distortions in the literature such as spatial variation of the fine structure constant~\cite{Sigurdson:2003pd} and vorticity from topological defects~\cite{Pogosian:2007gi}, we will not discuss their correspondence here as our objective here is to just point out that there are several cosmological distortions that are captured in Eqn.~(1). We do want to emphasize that not all the distortions in Eqn.~(1) correspond to cosmological distortions. For example as far as we know there are no known physical processes which can covert CMB temperature to polarization along the line-of-sight, hence there is no physical analog of temperature leakage distortions $(\gamma_1,\gamma_2),(d_1,d_2),$ and $q$. However we will now see that instrumental systematics can generate all the 11 distortion types, including leakage from temperature to polarization.

\subsection{ Instrumental Systematics}
Time varying errors produced by detector systems translate into position dependent distortions in the map. Even if the detector response is not time varying, for multiple detector systems if response of the different detectors have different systematics then that also results in position dependent distortion in the map. A map making process can be encapsulated as
\begin{eqnarray}
{\bf d}={\bf A s}+{\bn}\,,
\end{eqnarray}
here ${\bf d}$ is a vector containing time ordered data; ${\bf A}$ is a pointing matrix and encodes sky pixel location for each observation, scanning strategy, and beam effects; ${\bf s}$ is a map we want to reconstruct and ${\bn}$ is instrument noise. The relationship between the instrumental systematics and map distortion depends on scanning strategy. For a given experiment design and scanning strategy, starting from detector error models, one can calculate the distortions in the map.

Now we point out correspondence between 11 distortion fields in Eqn.~(1) and various instrumental effects (for details refer to~\cite{HHZ}). The effect of detector gain response miscalibration is captured by our modulation field $a({\bn})$. The effect of errors in the detector orientation is described by the $\rot({\bn})$ field. If the two polarization states are measured separately, then the differential gain is captured in our monopole leakage field $(\gamma_1({\bn}), \gamma_2({\bn}))$. Apart from the errors in the detectors there are also errors due to finite size of the beam. Let us consider an example of beam of the form  
\begin{align}
{\cal B}(\bn; {\bf b}, e) = {1 \over 2\pi \sigma^2 (1-e^2)}\exp \Big[
 -{1 \over 2\sigma^2}\Big( {(n_1-b_1)^2 \over (1+e)^2} +
{(n_2-b_2)^2 \over (1-e)^2} \Big) \Big]
		   \,,
\label{eqn:beammodel}
\end{align}
where ${\bf b}$ is the beam offset, $\sigma$ is the mean beam-width, and $e$ is the ellipticity. 
These parameters could be different for the different polarizations, and only the difference in the beams enters into the CMB measurements. Assuming any distortion of the beam are relatively small compare to typical scales of the beam on either direction, one can show the correspondence of beam distortions with the distortion parameter's considered in the paper
\begin{align}
\sigma {\bf p} &= ({\bf b}_a + {\bf b}_b)/2 \,, \nonumber\\
\sigma {\bf d} &= ({\bf b}_a - {\bf b}_b)/2 \,, \nonumber\\
e_s &= (e_a + e_b)/2  \,,\nonumber\\
q &= (e_a - e_b)/2 \,, 
\end{align}
Hence all the 11 distortion fields in Eqn.~(1) correspond to some sort of instrumental systematics~\cite{HHZ}.


\begin{figure*}[t]
\includegraphics[width=62mm,angle=-90]{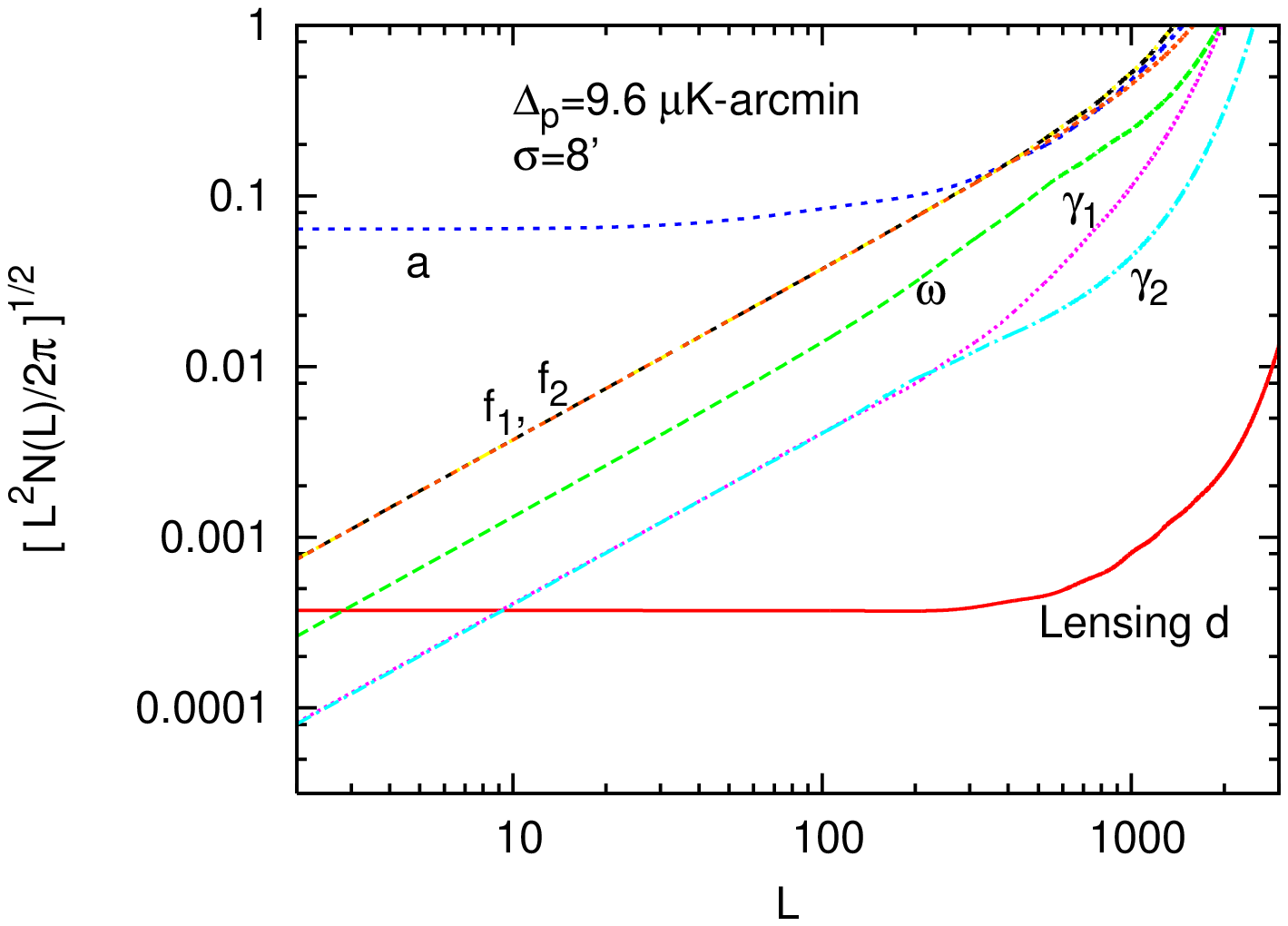}
\includegraphics[width=62mm,angle=-90]{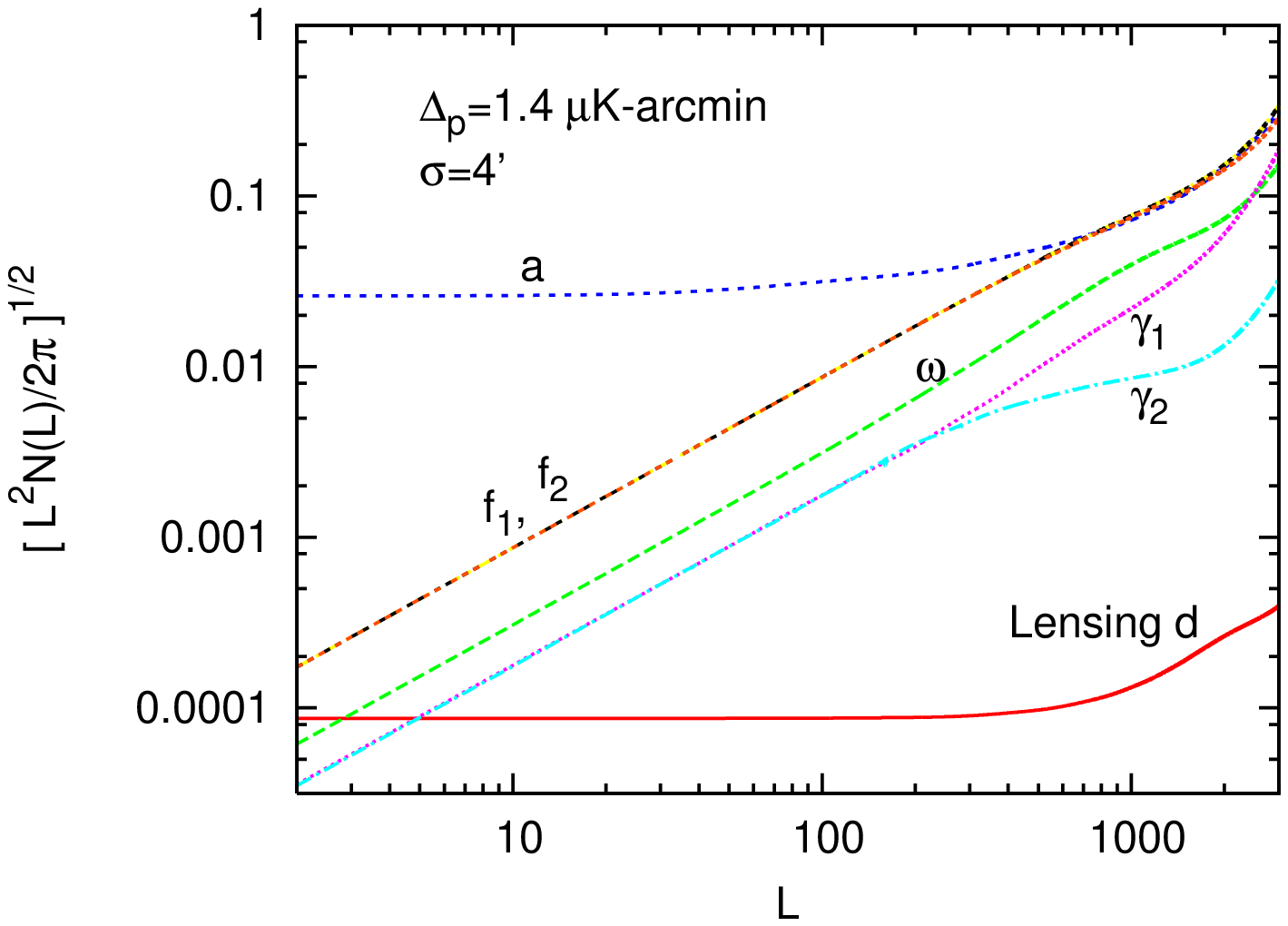}
\includegraphics[width=62mm,angle=-90]{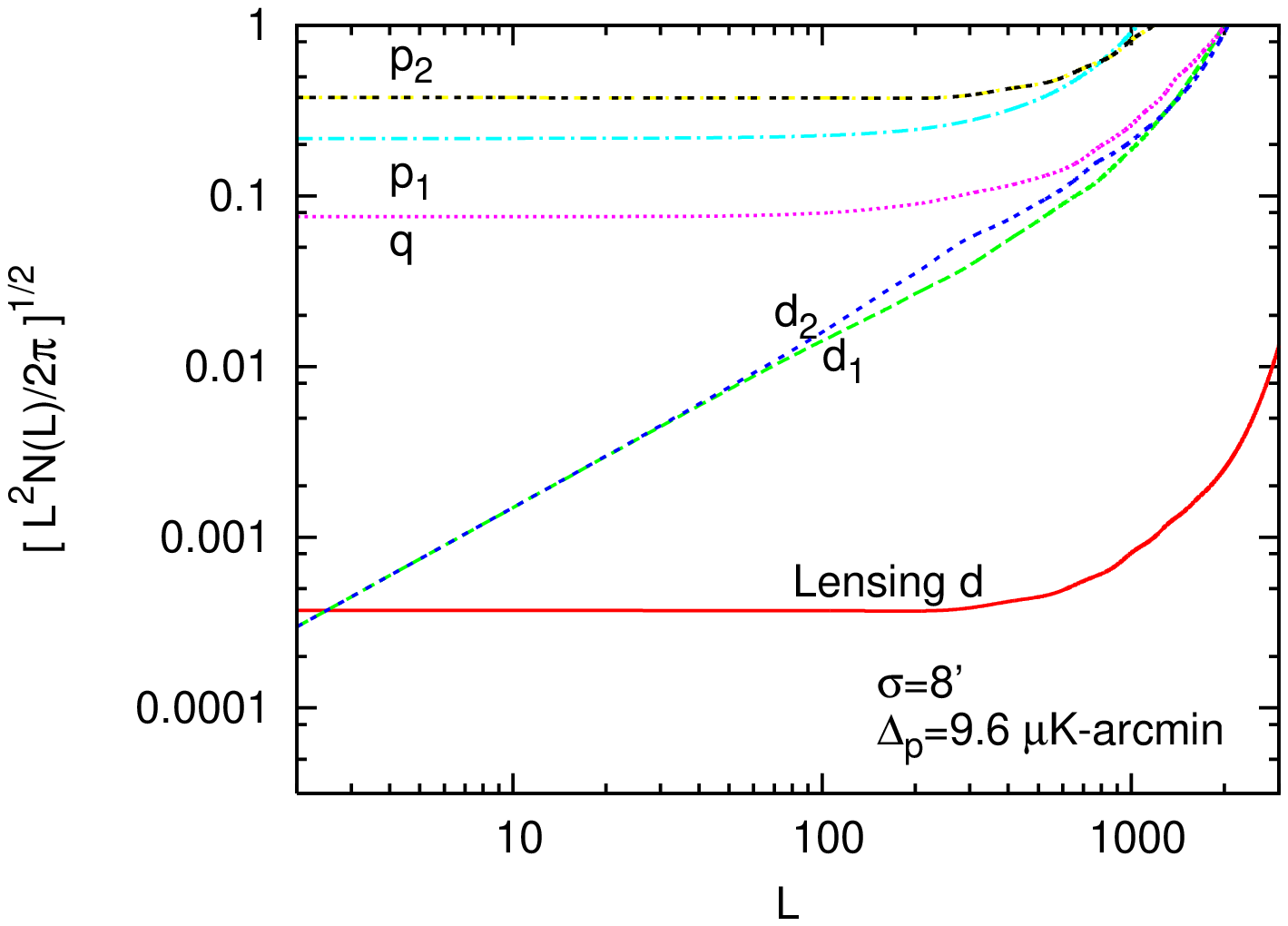}
\includegraphics[width=62mm,angle=-90]{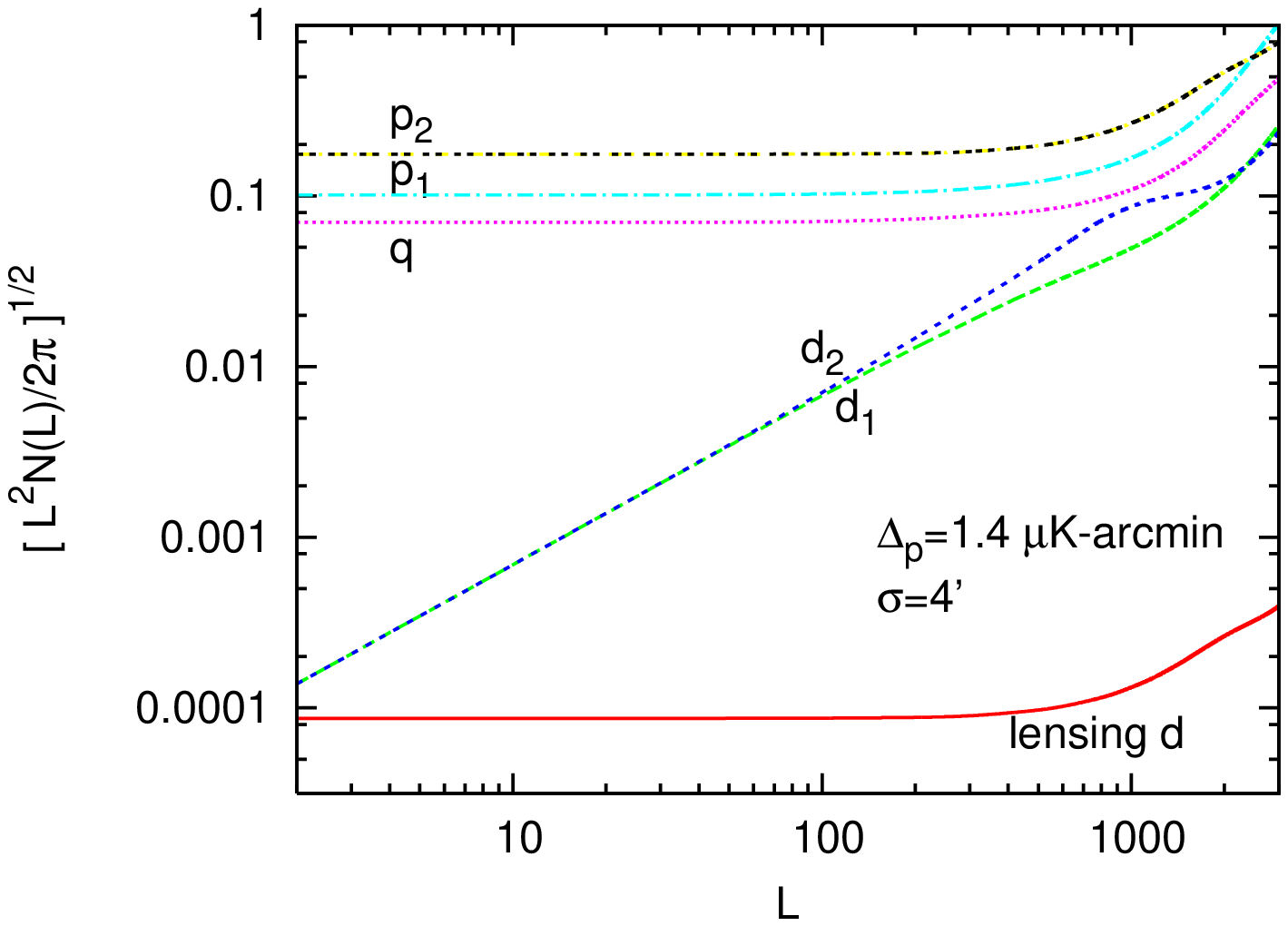}
\caption{Estimator Gaussian noise $N^{\cal D}_{EB}(L)$ as a function of multipole $L$ for the EB estimator. {\it Upper panels} show the noise for modulation $a$, rotation $\rot$, monopole leakage $(\gamma_1, \gamma_2)$, and spin-flip $(f_1,f_2)$; while the {\it lower panels} show the noise for deflection ${\bf p}$, dipole leakage $(d_1,d_2)$, and quadrupole leakage $q$. For the left panels we assume the instrument with polarization noise $\Delta_p=9.6\mu$K-arcmin and beam FWHM $=(8\ln2)^{1/2}\sigma=8'$. For the right panel we assume a CMBPol-like instrument with polarization noise $\Delta_p=1.4\mu$K-arcmin and beam FWHM $=(8\ln2)^{1/2}\sigma=4'$. Calibration $a$ is dimensionless, rotation $\rot$ is in radians, deflection $p$ and dipole leakage $(d_1,d_2)$ are in the units of Gaussian beam $\sigma$; and quadrupole leakage $q$ is in the units of $\sigma^2$. For reference we also show the Gaussian noise for lensing deflection estimator.}
\label{fig:N-type2}
\end{figure*}

\begin{figure*}
\includegraphics[width=62mm,angle=-90]{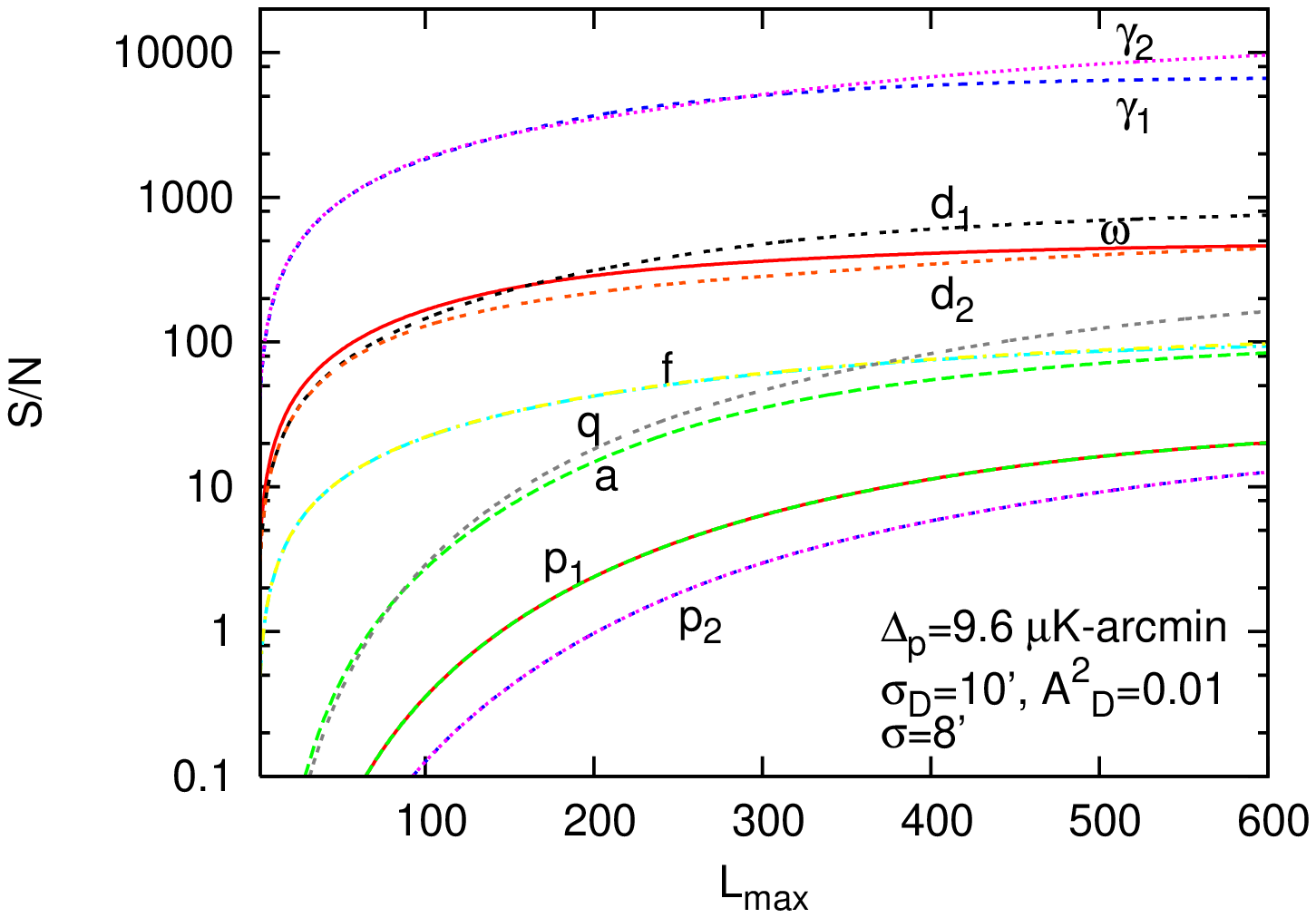}
\includegraphics[width=62mm,angle=-90]{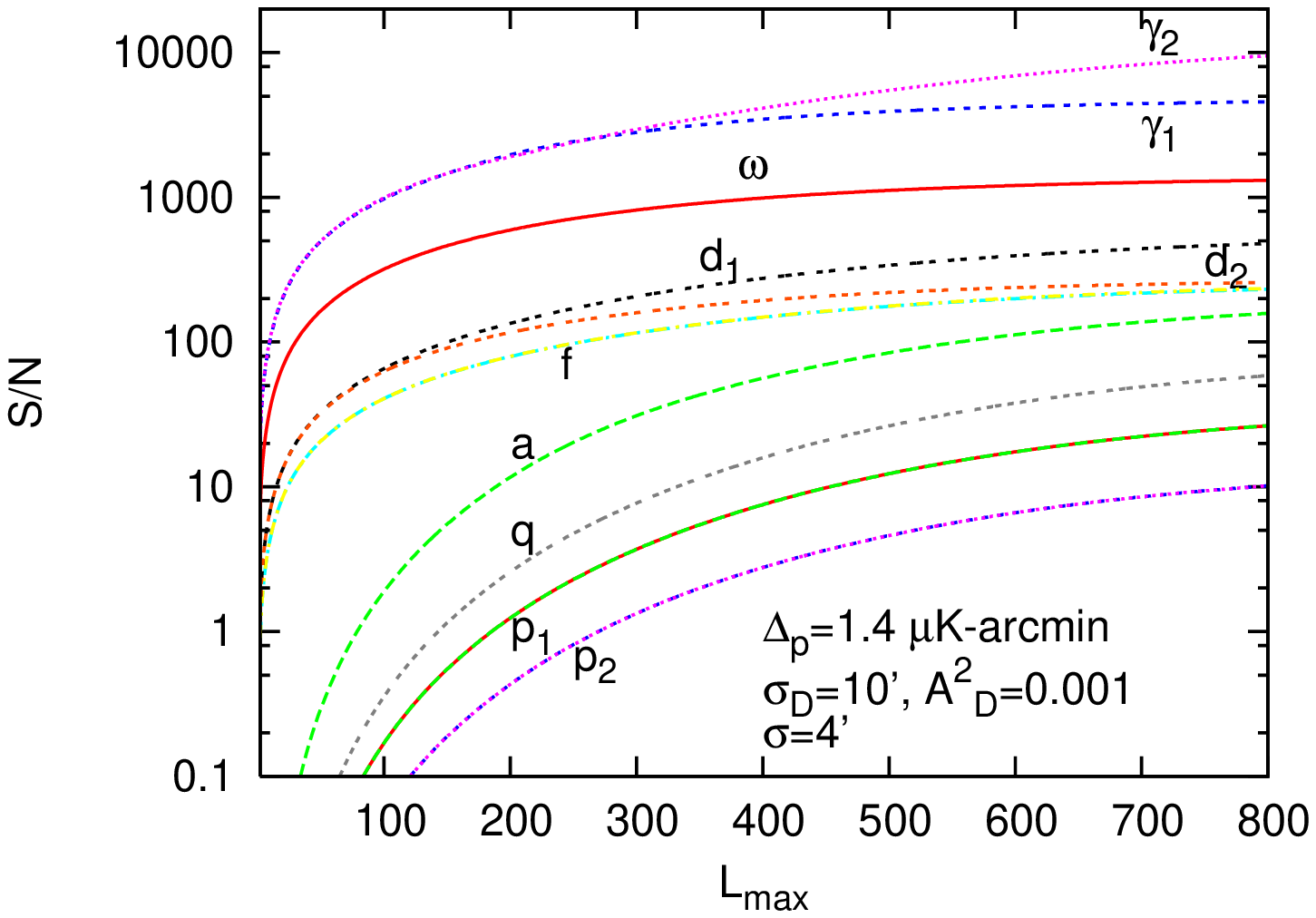}
\caption{Total signal-to-noise $S/N$ for the $EB$ estimator as a function of maximum multipole $L_{max}$. We assume that the distortion field power spectrum is white noise above a coherence scale $\sigma_{\cal D}$, with {\it rms} amplitude $A_{\cal D}$. The exact form of the distortion power spectrum is given by Eqn.~(\ref{eqn:css}). The left panel assumes the instrumental noise and beam properties of EXP-balloon experiment, while the right panel assumes that of CMBPol experiment. In both the panels we have assumes full-sky coverage.}
\label{fig:s2n}
\end{figure*}

\begin{table}
\begin{center}
\begin{tabular}{c||cc|cc||cc}
Column 1&  \multicolumn{4}{c||} {Column 2} & \multicolumn{2}{c} {Column 3}  \\ 
\hline 
 & \multicolumn{4}{c||} {$\Big( \frac{A^{\text{max}}_{\cal D}}{A^{\text{min}}_{\cal D}}\Big) \Big(\frac{f_{sky}}{f_\star}\Big)^{1/4}   \Big( \frac{r}{r_\star}\Big)^{1/2}$  } & \multicolumn{2}{c} {Maximum allowed {\it rms} $A^{\text{max}}_{\cal D}$}  \\ 
&  \multicolumn{4}{c||} {} & \multicolumn{2}{c} {for B-modes detection~\cite{HHZ}}  \\
\cline{2-7}
       Distortion Type    &  \multicolumn{2}{c|}{EXP-balloon $(f_\star=0.01, r_\star=0.05)$}  &  \multicolumn{2}{c||}{CMBPol $(f_\star=0.8, r_\star=0.005 )$} &  \multicolumn{2}{c}{$\Big(\frac{r}{0.005}\Big)^{1/2}$} \\ 
          & $\sigma_s=10'$ & $\sigma_s=120'$ &$\sigma_s=10'$ & $\sigma_s=120'$  & $\sigma_s=10'$ & $\sigma_s=120'$\\ \hline
Rotation $\rot$             &3.4 & 11.9 & 16.72 &  49.02 & 0.015 & 0.011 \\                  
Modulation $a$              &6.0 & 5.13 & 25.46 &  12.73 & 0.06  &  0.049 \\
Monopole leakage $\gamma_1$ &1.9 & 2.13 & 4.75 &  4.65 & 0.0023 &  0.0006\\
Dipole leakage $\gamma_2$   &2.0 & 1.7 & 6.46 &   3.9 & 0.0019 &  0.0005 \\
Spin flip $f_1$             &6.2 & 17.9 & 29.35 &  73.15 & 0.061 & 0.046\\
Spin flip $f_2$             &6.3 & 17.6 & 28.7 & 71.53 & 0.059  & 0.045\\
Dipole leakage $d_1$        &2.2 & 5.23 & 5.4 &  10.54 & 0.0077 &   0.0053\\
Dipole leakage $d_2$        &1.7 & 5.38 & 3.8 &  11.11 & 0.0077 &   0.0056\\
Quadrupole $q$              &1.8 & 4.1 & 3.32 &  3.55 & 0.0124  &   0.0394\\
Deflection $p_1$            &38.2 & 19.4 & 132.7 & 40.3 & 0.75   & 0.53 \\
Deflection $p_2$            &4.4  & 15.5 & 10.8 &  24.8 & 0.098 & 0.57 \\
\hline
\end{tabular}
\end{center}
\caption{Here we show that using the EB correlations, the distortions will be detected at high significance before the distortions show up as B-modes. In column 1 we show the distortion type in consideration. For a given distortion type, the numbers in column 3 show the maximum allowed distortion {\it rms} $A^{\it max}_{\cal D}$ above which the distortions will be detected as B-modes with an experiment which is sensitive to $r=0.005$. These numbers are from~\cite{HHZ}. In column 2 we compare the minimum detectable distortion {\it rms} amplitude $A^{\text{min}}_{\cal D}$ (for experiments EXP-balloon and CMBPol) using our EB estimator with the maximum allowed distortion {\it rms} $A^{\text{max}}_{\cal D}$ presented in column 3. The minimum detectable {\it rms} $A^{\text min}_{\cal D}$ was obtained using Eqn.~(\ref{eqn:s2n}) and seeking the value of $A_{{\cal D}}$ for which total signal-to-noise becomes unity. We have considered two choices of coherence length $\sigma_{\cal D}=10'$ and $\sigma_{\cal D}=120'$. For same $\sigma_{\cal D}$, the numbers in column 2 are much larger than the numbers in column 3, this implies that the distortions will be detected using $EB$ estimator before they will be detected as B-mode power spectrum. }
\label{tab:comparewithHHZ}
\end{table}

\section{Results and Discussion}
\label{sec:results}
Here we discuss the prospects of using our estimators for detecting the distortions in the CMB. We study the estimator variance for two experimental setups, (1) experiment with noise sensitivity of $\Delta_p=9.6 \mu$K-arcmin and beam FWHM $\sigma=8'$, typical of upcoming ground and balloon based CMB polarization experiments (EXP-balloon hereafter) and (2) a futuristic CMBPol like satellite experiment with noise sensitivity $\Delta_p=1.4 \mu$K-arcmin and beam FWHM $\sigma=4'$ (CMBPol hereafter).  We choose a standard
fiducial model with a flat $\Lambda CDM$ cosmology, with parameters
described by the best fit to WMAP5~\cite{Komatsu2009}, given by
$\Omega_b=0.045, \Omega_c=0.23, H_0=70.5, n_s=0.96, n_t=0.0,$ and
$\tau=0.08$. We calculate the theoretical lensed and
unlensed CMB power spectrum from publicly available code CAMB\footnote{http://camb.info/}~\cite{Lewis:1999bs}. We assume zero distortions as our fiducial model, except for the lensing of CMB by the large-scale structure whose effect we include in the variance by using the lensed CMB power spectrum.  For the experiments in consideration, the EB estimator is most sensitive so we will restrict our discussion to that.  

In Fig. (\ref{fig:N-type2}) we show the Gaussian noise $N^{\cal D}_{EB}(L)$ for all the 11 distortion fields. Notice that noise $N^{{\cal D}}_{EB}(L)$ for distortions $a,q,$ and $(p_1,p_2)$ whose filter $f_{EB}(\bfl_1,\bfl_2)$ contains $\sin(\varphi_{\bfl_1}-\varphi_{\bfl_2})$ scales as $N(L)\propto 1/L^2$ for low $L$. This scaling can be understood by noting that $N(L)\propto \frac{1}{f^2{(\bfl_1,\bfl_2)}}$ and for small $L$, $l_1\approx l_2$ and hence $\sin(\varphi_{\bfl_1},\varphi_{\bfl_2})\approx \varphi_{\bfl_1 \bfl_2}\propto L/l_1$. In a similar way the scaling of other distortions noise can be understood, for example noise $N(L)$ for rotation $\rot$ whose filter $f_{EB}(\bfl_1,\bfl_2)$ contains $\cos(\varphi_{l_1},\varphi_{l_2})$ is roughly constant $N(L)\approx constant$ at low $L$. This is because for small $L$, $l_1\approx l_2$ and hence $\cos(\varphi_{\bfl_1},\varphi_{\bfl_2})\approx 1$.  

For a given multipole $\ell$, if the distortion power spectrum is larger than the Gaussian noise $N^{\cal D}(\ell)$ then the estimator can reconstruct the distortion map with $(S/N)_\ell > 1$. However if the distortion power spectrum is smaller than the Gaussian noise than the distortion map cannot be reconstructed but one can still statistically detect and constrain its size. Later in the section we consider an application of the estimator for doing diagnostics for the primordial B-modes. There we will quantify the level of distortions that can be detected using our estimators, and how they compare to other methods.

In Fig.~\ref{fig:s2n} we show total signal-to-noise for all the 11 distortions as a function of $\ell_{max}$ for EXP-balloon and CMBPol experiments. To calculate the $S/N$ we have assumed that the power spectrum of distortion is given by Eqn.~(\ref{eqn:css}). The $S/N$ scales as $A^2_{\cal D}$. In the figure we have chosen {\it rms} $A_{\cal D}=\sqrt{0.01}$ for EXP-balloon experiment and $A_{\cal D}=\sqrt{0.001}$ for CMBPol experiment\footnote{The values of {\it rms} in Fig.~\ref{fig:s2n} were mostly chosen for visual purpose. These {\it rms} values roughly correspond to $10~\sigma$ detection of deflection distortion $p_2$.}. The coherence scale is taken to be $\sigma_{\cal D}=10'$ for both the experiments. For same distortion {\it rms}, among all the distortions,  monopole leakage has the highest $S/N$. This is because the amplitude of temperature fluctuations is much larger than the polarization amplitudes and monopole leakage represent a distortion where temperatures gets converted into polarization.

\begin{figure*}[t]
\includegraphics[width=110mm,angle=0]{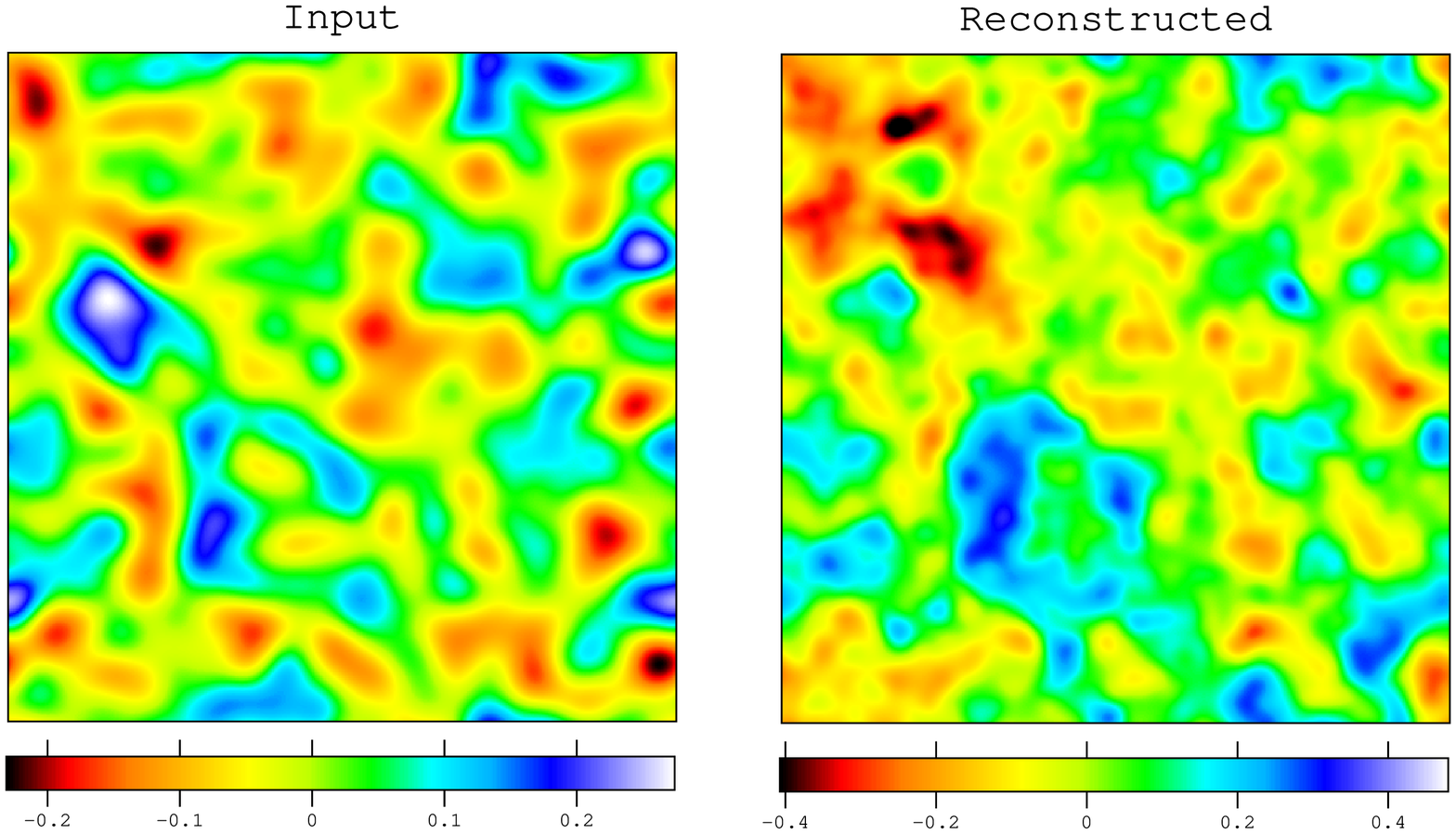}
\includegraphics[width=110mm,angle=0]{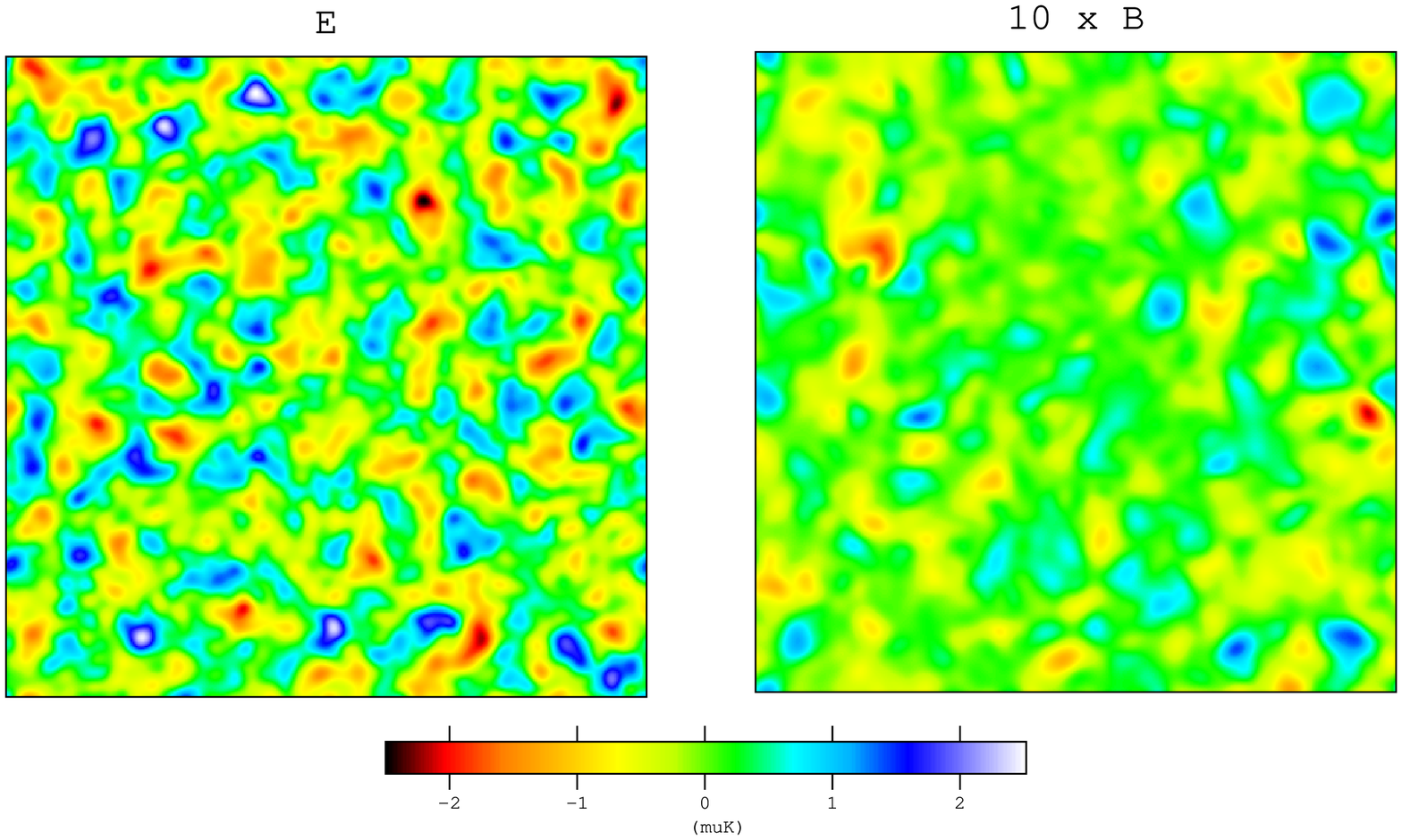}
\includegraphics[width=62mm,angle=-90]{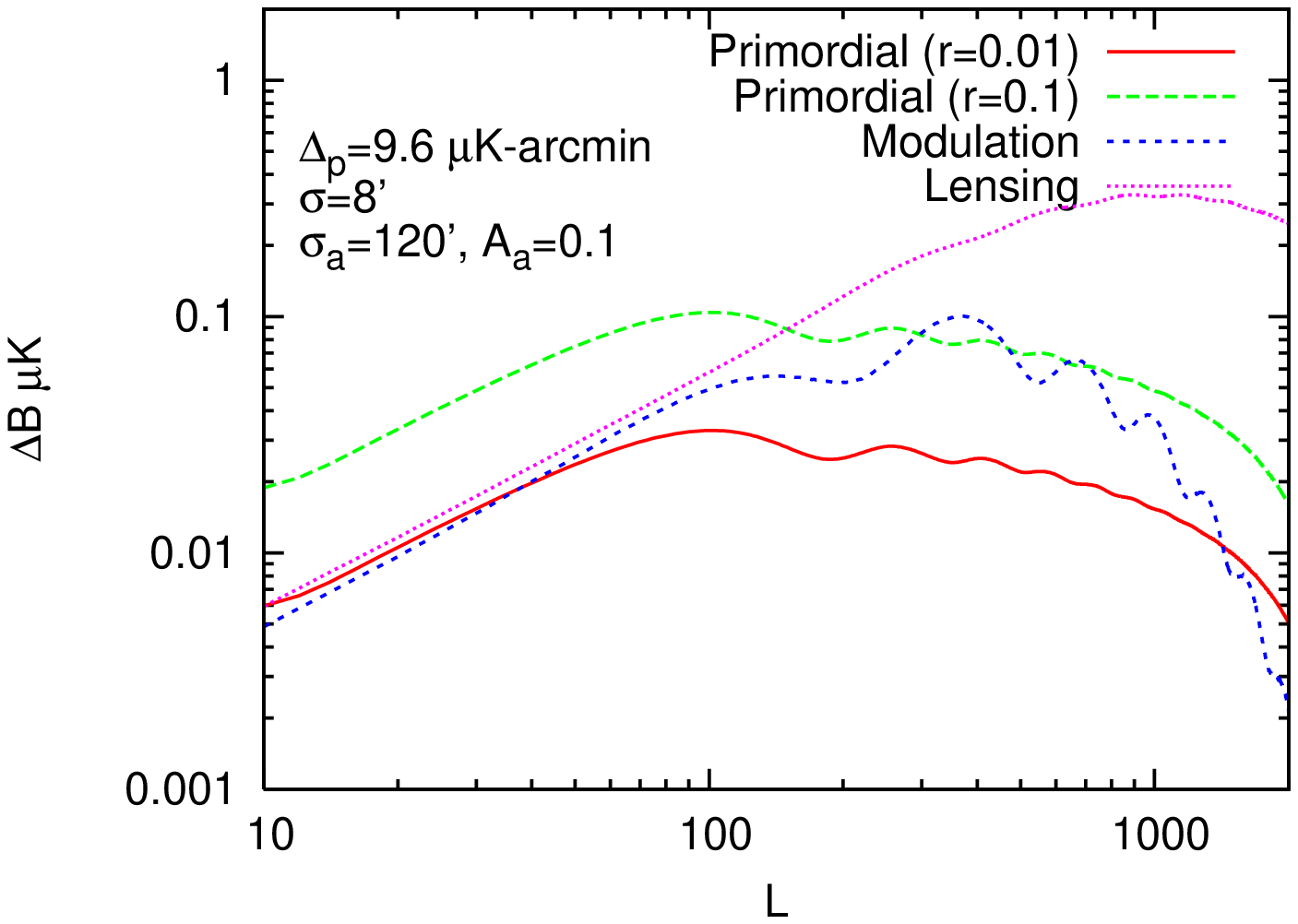}
\includegraphics[width=62mm,angle=-90]{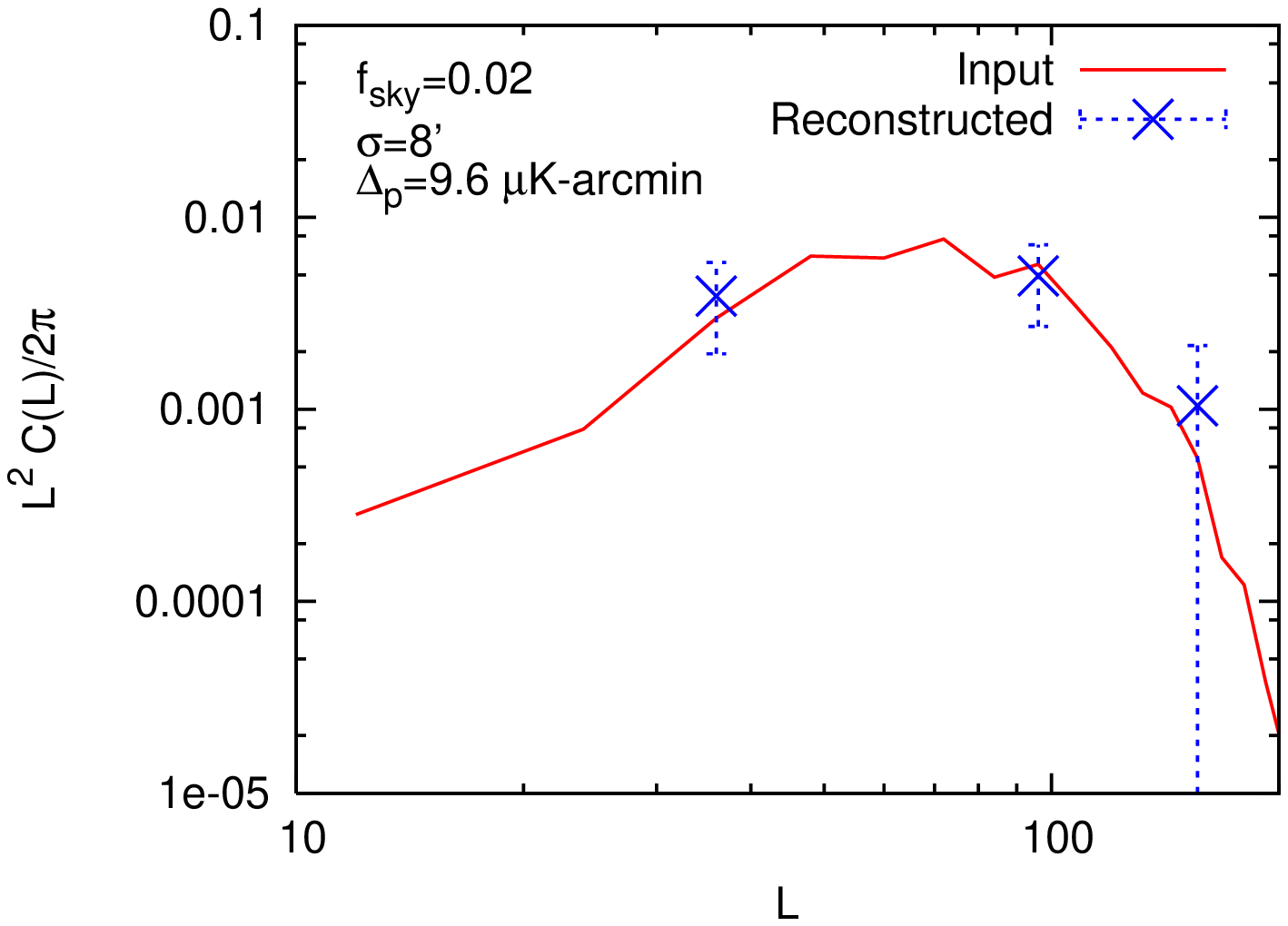}
\caption{The upper left map shows the input modulation map $a^{{\text input}}(\bn)$, the upper right map shows the reconstructed $\hat a(\bn)$ map, the middle left map shows the E map after primordial E map has been distorted via input modulation map (upper left map), the middle right map shows the B-polarization field generated by input modulation. In our simulations we have assumed primordial B-modes to be zero. The lower left map compares the B-modes generated by modulation with the primordial B-modes (for $r=0.1$, $r=0.01$) and lensing induced B-modes. The lower right panel compares the power spectrum of input modulation field with the reconstructed binned power spectrum. The size of the simulation patch was $30\times 30$ square degree. The instrumental noise $\Delta_p=9.6 \mu$K-arcmin, and beam FWHM $\sigma=8'$ correspond to EXP-balloon, an experiment representative of upcoming balloon and ground based experiments. }
\label{fig:one_realization_simulation}
\end{figure*}

Although the estimators are unbiased when averaged over CMB realizations, for a given CMB realization (which is the case in reality) the reconstruction is noisy even for the cosmic variance limited experiment with perfect beam. In Fig. (\ref{fig:one_realization_simulation}) we show the reconstruction of distortion for a given CMB realization. We consider a modulation distortion and show the reconstruction ability for EXP-balloon like experiment. The simulation patch size is $30\times 30$ square degrees. We assume the modulation to be a Gaussian field with the underlying power spectrum given by Eqn.~(\ref{eqn:css}). We have chosen the coherence $\sigma_{{\cal D}}$ and {\it rms} $A_{{\cal D}}$ in such a way that the modulation induced B-modes resemble the level and shape of primordial B-modes that EXP-balloon is sensitive too. In the lower left panel we compare the B-modes generated from the modulation field with the primordial B-modes with $r=0.01$ and $r=0.1$; and lensing induced B-modes. The upper left map shows the realization of modulation field $a^{{\text input}}(\bn)$ that we use for the simulation. The upper right panel shows the reconstructed modulation field for a single CMB realization. It is clear that the reconstruction is noisy however the reconstructed modulation traces the features of input modulation. We have smoothed the input and reconstructed maps with Gaussian beam FWHM of $1^\circ$ so that it is easy to see the common features.  The middle panel shows the CMB E and B fields which were used to reconstruct the modulation field. The middle left map shows the E-polarization of the CMB after being distorted using input modulation field $a^{{\text input}}(\bn)$. The middle right map shows the B-mode polarization generated by the modulation field $a^{{\text input}}(\bn)$, note that these B-modes are solely due modulation of primordial E field as we have assumed zero primordial B-modes for our fiducial model. In the lower right figure we compare the reconstructed binned power spectrum with the power spectrum of input modulation $a^{{\text input}}(\bn)$.

\begin{figure*}[t]
\includegraphics[width=62mm,angle=-90]{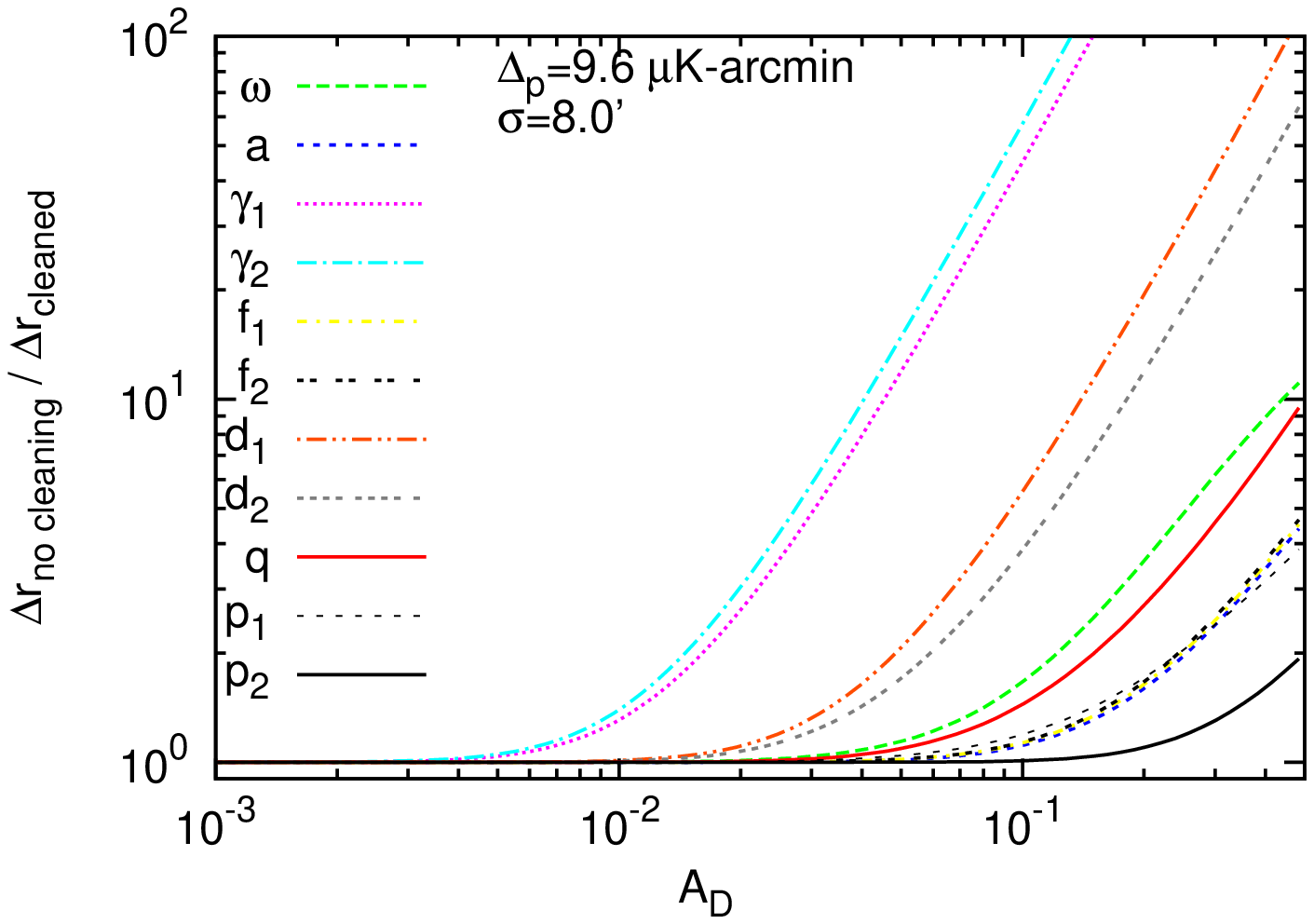}
\includegraphics[width=62mm,angle=-90]{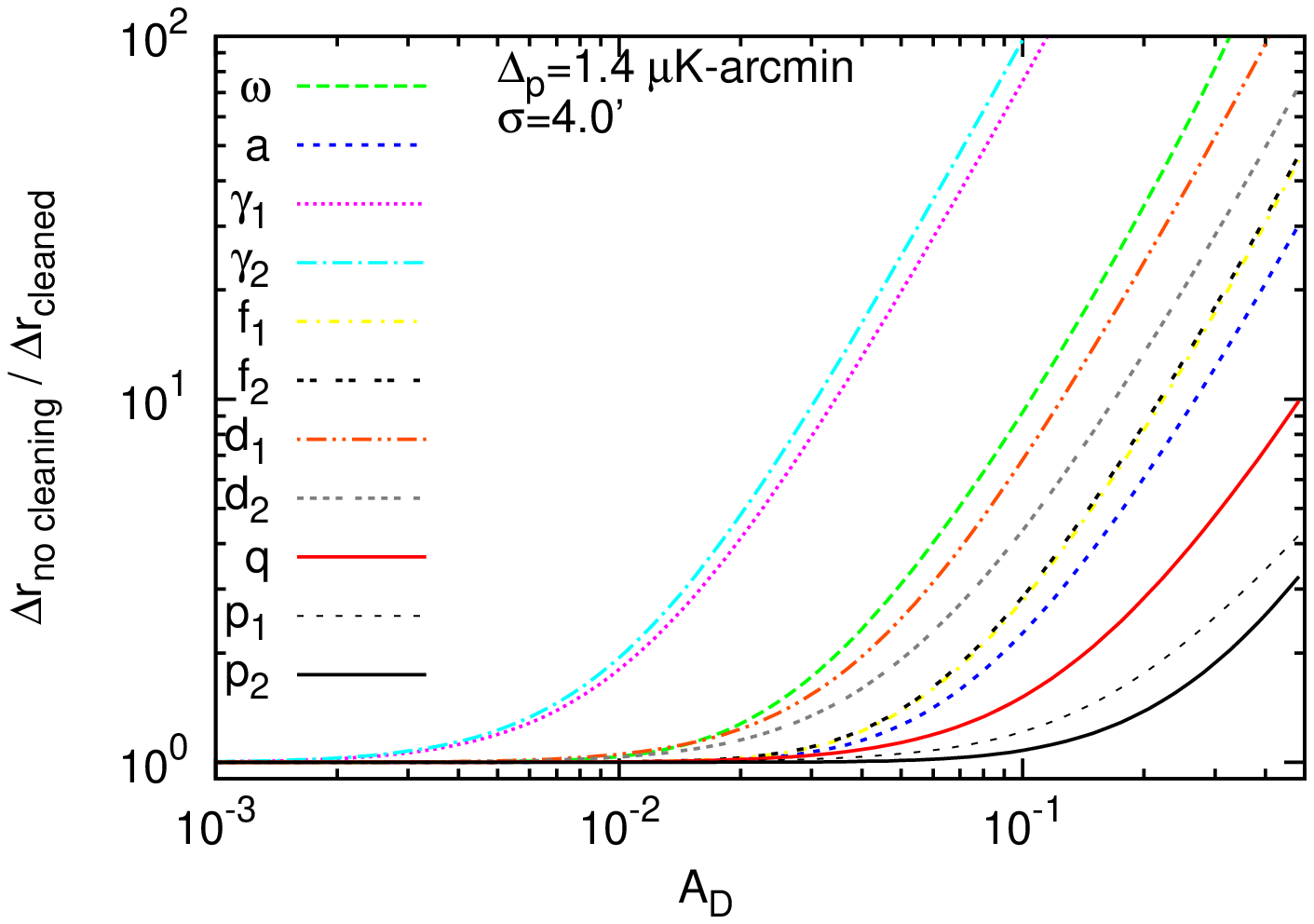}
\caption{Ratio of tensor-to-scalar uncertainty $\Delta r$ without and with cleaning as a function of distortion {\it rms} $A_{\cal D}$. We have assumed convergence length to be $\sigma_{\cal D}=10'$. The left panel is for the EXP-balloon experiment, and right panel is for the CMBPol experiment. The ratio of uncertainty is independent of tensor perturbations, and sky fraction.}
\label{fig:t2s}
\end{figure*}

\subsubsection*{Primordial B-mode Diagnostics} As we discussed in the introduction, detection of primordial gravitational wave provides information about the early universe. So far primordial gravitational waves have not been detected, there are only upper limits on their amplitude, see~\cite{ligo-gwave}, for a current observational bounds on the level for primordial gravitational waves. 
If the B-modes are observed what diagnostics test can be done to make sure that they are primordial~\cite{Baumann:2009mq}? As we have already shown that many distortions in the CMB which produces B-modes also produce off-diagonal correlations in the CMB. Using the off-diagonal correlations in the CMB, we have also constructed estimator to reconstruct the distortions. Now we show that our estimators are more sensitive to distortion than the B-modes and any non-primordial B-modes which can be detected, can be detected using our estimators at much higher significance.

In Table~\ref{tab:comparewithHHZ} we compare the minimum detectable level of distortion {\it rms} $A^{\text min}_{{\cal D}}$ using our $EB$ estimator with the maximum allowed distortion {\it rms} for primordial B-modes detection. We consider two experiments EXP-balloon and CMBPol, which are sensitive to scalar-to-tensor ratio of $r=0.05$ and $r=0.01$, respectively. We find that the $EB$ estimator will be able to see any spurious contamination before it gets detected as B-modes. In column 1 we tell the distortion type in consideration. For each distortion type, the column 3 shows the maximum distortion {\it rms} $A^{\text{max}}_{\cal D}$ above which the distortion will be detected as B-modes for an experiment that is sensitive to $r=0.005$. In column 2 we show the ratio of maximum allowed {\it rms} given in column 3 and the minimum detectable {\it rms} using the our EB estimator. For EXP-balloon, which is typical of upcoming balloon based experiment, we assume sky fraction to be $1\%$. For CMBPol experiment we assume full sky fraction to be $f_{sky}=0.8$. For distortion field we have considered two choices of coherence length $\sigma_{\cal D}$. For a given coherence $\sigma_{\cal D}$, the fact that numbers in column 2 are larger than unity says that the distortions will be detected at higher significance using EB correlations than using B-mode power spectrum of the CMB.

Detection of distortion ${\cal D}$ through these distortion estimators clearly determines the presence of non-primordial B-modes. In addition it may mean two things 1) either the source of the distortion is instrumental systematics or 2) the distortion signal is cosmological (although not primordial\footnote{Primordial fluctuations do not generate $\langle EB \rangle$ correlations, and hence our quadratic estimator will not see anything.}). If the source of the detection is instrumental systematics, the significance of the detection will drop as the instrumental systematics is controlled better (or another instrument with better systematics (corresponding to distortion ${\cal D}$) control will not see the detection using the ${\cal D}$ distortion estimator\footnote{The Distortions due to systematics will not correlate between different experiments.}). However if the systematics is controlled well below the noise level $N^{{\cal D}}(L)$ of ${\cal D}$ distortion estimator, and the signal is still present, then the detection should be attributed to cosmological signal. All cosmological distortions in the CMB probe interesting cosmological processes and in some cases probe physics beyond the standard model~\cite{Yadav_etal2009}. Hence detection of cosmological distortion will be extremely exciting. 

So far we have only shown that the distortions can be detected in the power spectrum. However if the $S/N$  per mode is much larger than unity, our estimators can be used to reconstruct the maps of the distortion fields. The reconstructed distortion then can be used to clean the CMB B-modes by subtracting the spurious B-modes from the observed B-modes, hence improving the sensitivity for the tensor B-modes. Fig.~\ref{fig:t2s} demonstrates the improvement in the uncertainty of tensor-to-scalar ratio $\Delta r$ due to the cleaning. We show the ratio of tensor-to-scalar uncertainty $\Delta r$ without and with cleaning as a function of the distortion {\it rms} $A_{\cal D}$. In Appendix~B we describe the B-modes cleaning process. While both  $\Delta r_{\text{with cleaning}}$ and $\Delta r_{\text{no cleaning}}$ depend on the value of tensor B-mode power spectrum and sky fraction $f_{sky}$, the ratio is independent of them. It is clear from the figure that larger the {\it rms} of the distortions, larger the reduction in the error of tensor-to-scalar $\Delta r$ due to cleaning~\footnote{Note that if the {\it rms} is below the requirement from tensor B-mode detection (i.e. column 3 of Table II), one does not reduce the uncertainty in $r$ by much from the cleaning process, for example for CMBPol. For example for CMBPol the the uncertainty $\Delta r$ reduces by only factor of $\sim 5$ for the deflection $p_1$ and is even smaller for other distortions.}.

\section{Summary}
\label{sec:summary}
We have considered general line-of-sight distortions in the CMB. These distortions can be described by $11$ fields and their effect on CMB polarization is given by Eqn.~(1). We show that these fields correspond to various known instrumental systematics and cosmological signals. Example of cosmological signals include rotation of plane of polarization due to magnetic fields or parity violating physics, screening effects of the patchy reionization, lensing of CMB by the large-scale structure. Instrumental systematics include gain fluctuation, differential gain, pointing, beam ellipticity, differential beam ellipticity, detector rotation.  All these distortions generate B-modes of CMB and also generate unique off-diagonal correlations in the $E$ and $B$ fields. The way the different modes of $E$ and $B$ get correlated depends on the distorting field $\cal D$ and can be compactly written as $\langle E(\bfl_1)B(\bfl_2)\rangle=f^{\cal D}(\bfl_1,\bfl_2) D(\bfl_1+\bfl_2)$. We construct quadratic estimators for all the $11$ distortion fields which utilizes the unique correlation signature in the observed CMB fields to reconstruct the distortion fields. We compute the variance of the estimators and show the level of distortions that can be detected in upcoming and future experiments such as CMBPol. As an example of our estimator we show that our estimator can detect the distortions before they can be detected as spurious B-modes in the power spectrum. This is particularly important diagnostics for primordial B-modes, since  distortions induced B-modes act as contamination for primordial B-modes detection.  Since there are only 11 distortions, it is worthwhile to look for them in CMB as systematics check or if instrumental systematics is well controlled, to look for cosmological signals.

\acknowledgments{A.P.S.Y. acknowledges support from IBM Einstein fellowship. M.Z. is supported by the David and Lucile Packard, the Alfred P. Sloan, the John D. and Catherine T. MacArthur Foundations, NSF grants AST-0506556, AST-0907969, and  PHY-0855425 and NASA NN-6056J406. We acknowledge stimulating conversations with Matthew McQuinn. We especially thank Daniel Baumann and Rahul Biswas for reading the manuscript and providing useful feedback.}



\appendix
\section{Estimating multiple Distortions simultaneously}                              
In the main text we discussed estimator for different distortion fields, while estimating them one at a time i.e. in the absence of any other distorting field. In this appendix we discuss how can we estimate multiple distortion fields simultaneously. For simplicity we will restrict the discussion to the EB estimator, the generalization to include other CMB fields is straight forward. To generalize our $EB$ estimator to multiple field we start by defining a $N$-by-$N$ matrix $F^{{\cal D}{\cal D}'}_\ell$ for each multipole $\ell$,
\ba
F^{{\cal D}{\cal D}'}_\ell &=& \intl{1}f^{{\cal D}}_{EB}(\ell_1,\ell_2)
({\bf C}^{-1})^{EE}_{\ell_1}f^{{\cal D}'}_{EB}(\ell_1,\ell_2)
({\bf C}^{-1})^{BB}_{\ell_2} \,, 
\ea
where ${\cal D}$ and ${\cal D}'$ run over  all the 11 distortions and lensing, $\{a,\rot, \gamma_1, \gamma_2, f_1, f_2, d_1, d_2, q, p_1, p_2, \phi\}$. An estimator in the presence of single distorting field was unbiased, 
\begin{eqnarray}
\big\langle\hat {\cal D}(\bfL)\big\rangle_{CMB}={\cal D}(\bfL)\,.
\label{eqn:unbiased}
\end{eqnarray} 
In the presence of multiple distorting fields the estimator as constructed in Eqn.~(10), may not be unbiased. In general the Eqn.~(\ref{eqn:unbiased}) gets modified as
\begin{equation} 
\langle \hat {\cal D}(L)\rangle_{CMB} = {\cal D}(L) + \frac{ \sum_{{\cal D}'} F_L^{{\cal D}{\cal D}'} {\cal D}'(L)}{F^{{\cal D}{\cal D}}_L} \,,
\label{fisher_bias}
\end{equation}
where $\hat {\cal D}$ is an estimate of ${\cal D}$ in the presence of multipole distortions effect in the CMB. The level of bias depends on the off-diagonal Fisher matrix $F_L^{{\cal D}{\cal D}'}$ and the amplitude of distortion fields ${\cal D'}$. Eqn.~(\ref{fisher_bias}) can be inverted iteratively to get an unbiased estimate of a field ${\cal D}$. The number of iterations required to converge depends on the signal levels and correlation between the fields. 
The correlation between two fields at a give multipole $L$ can be calculated as
\begin{equation}
{\cal C}_{{\cal D}{\cal D}'}=\frac{F^{{\cal D}{\cal D}'}_\ell}{\sqrt{F^{{\cal D}{\cal D}}_\ell F^{{\cal D'}{\cal D}'}_\ell}}
\end{equation}
In Table~\ref{tab:fishermatrix} we give the Fisher matrix at $L=100$ for EXP-balloon experiment. Most of the distortion fields are uncorrelated\footnote{As we have shown in Sec.~\ref{sec:correspondence}, lensing deflection is same as deflection $p_2$. Hence as expected the correlation coefficient between lensing and $p_2$ is unity, ${\cal C}_{p_2 \text{lens.} d}=1$.}, however there are some fields which are highly correlated. For example the largest correlation is between quadrupole leakage $q$ and dipole leakage $d_2$, with the correlation coefficient ${\cal C}_{q d_1}=.9$. The second largest correlation is between deflection $p_1$ and rotation $\rot$, with the correlation coefficient ${\cal C}_{p_1\rot}=.8$. The next three largest are ${\cal C}_{p_2a}=-0.2$, ${\cal C}_{q \gamma_1}=.15$ and ${\cal C}_{a \gamma_1}=.14$.

The variance of the distortion ${\cal D}$ will be given by $\big(F^{-1}\big)_{{\cal D}{\cal D}}$. Multiple field formalism for other estimators $(TT,TE,EE,TB)$ and the minimum variance estimator is a straight forward generalization of the $EB$ estimator we discuss here.

\begin{table*}
\begin{center}
\begin{tabular}{c|cccccccccccc|}
\hline          & Lens. $d$  &  $\rot$    &   $a$   & $\gamma_1$     & $\gamma_2$   &  $f_1$ & $f_2$ & $d_1$ & $d_2$ & $q$ & $p_1$ & $p_2$      \\  \hline
Lens. $d$       & 0.1E+15  &-0.2E-06 &0.1E+10  &0.5E+09  &-0.9E-07  &-0.2E+08    &0.6E-07   &0.2E-07   &0.4E+09  &0.8E+08  &0.4E-08  &-0.1E+10 \\
     $\rot$     &&0.8E+07   &-0.1E-10 &-0.7E-10 &-0.1E+07  & 0.2E-10    &0.3E+03   &-0.2E+06  &-0.4E-11  &-0.3E-11  &0.4E+06   &0.2E-11 \\
     $a$        &&&0.2E+06  & 0.6E+06 &-0.8E-11  &0.1E+04     & -0.2E-11 &0.7E-11   &-0.1E+05  &0.4E+04   &-0.1E-11  &-0.1E+05 \\
     $\gamma_1$ &&&&0.9E+08  &-0.2E-09  &0.1E+05     &0.3E-10   &0.6E-10   &-0.6E+06  &0.8E+06   &0.3E-13   &-0.5E+04 \\
     $\gamma_2$ &&&&&0.9E+08   &0.1E-11  &-0.5E+04  &0.5E+06   &0.9E-12   &-0.5E-10  &-0.6E+04  &0.3E-11 \\
     $f_1$      &&&&&&0.1E+07   &0.7E-12   &-0.7E-11  &-0.2E+04  &-0.2E+03  &0.7E-12   &0.2E+03 \\
     $f_2$      &&&&&&&0.1E+07   &0.2E+04   &0.2E-10   &-0.2E-11  &0.1E+03   &0.7E-12 \\
     $d_1$      &&&&&&&&0.8E+07   &0.4E-10   &-0.3E-11  &0.5E+02   &-0.1E-12 \\
     $d_2$      &&&&&&&&&0.6E+07   &0.1E+07   &0.3E-11   &-0.4E+04 \\
     $q$        &&&&&&&&&&0.2E+06   &-0.1E-12  &-0.8E+03 \\
     $p_1$      &&&&&&&&&&&0.3E+05   &-0.6E-13 \\
     $p_2$      &&&&&&&&&&&&0.1E+05 \\
\hline
\end{tabular}
\end{center}
\caption{Fisher matrix $F^{{\cal D}{\cal D}'}_{L}$ at $\ell=100$ for EXP-balloon experiment with $\Delta_p=9.6 \mu$K-arcmin and beam FWHM$=8'$.}
\label{tab:fishermatrix}
\end{table*}

\section{Self calibrating the CMB} In the main text we have only shown that the non-primordial contribution to the B-modes can be detected using the EB estimators. Here we turn to the question that whether we can clean the CMB from the distortions and hence improve the sensitivity to primordial B-modes. It is also useful to correct the systematics induced distortions as they also affect lensing reconstruction from CMB~\cite{SYZ09,Miller:2008zi}. If the estimator $S/N$ per mode is greater than unity, then the estimators can be used to reconstruct the map of the distortion fields without any assumptions about the statistics of the distortion fields (i.e. The distortion field does not have to be Gaussian). Once the distortion fields are reconstructed one can clean the observed CMB by un-doing the distortions in CMB. The process is similar to de-lensing~\cite{Hu:2000ee,Hirata:2003ka,Hirata:2002jy,Smith:2008an}. 

In principle one can clean the CMB field from distortions iteratively. For each iteration, the first step is estimating the distortion filed $\cal \hat D$ using the observed CMB $E$ and $B$ fields, i.e. using Eqn.~\ref{eqn:estimator}. Then constructing an estimate of distortion induced B-modes $\hat B(\bfl)$ given the distortion field $\cal D$, and observed CMB field $X$ can be calculated as
\begin{eqnarray}
\hat B(\bfl) = \int \frac{d^2\bfl_1}{(2\pi)^2} W^{\cal D}_{Y}(\bfl_1,\bfl_2) \frac{C^{XX}_{\ell_1} X(\bfl_1)}{C^{XX}_{\ell_1}+{\cal N}^{XX}_{\ell_1}}\frac{C^{\cal D \cal D}_{\ell_1} {\cal {\hat D}}(\bfl_1)}{C^{\cal D \cal D}_{\ell_2}+N^{\cal D \cal D}_{\ell_2}}\,,
\end{eqnarray}
where $X=E,\, Y=B$ for ${\cal D}={a,\rot,(f_1,f_2), (p_1,p_2)}$ and $X=T,\, Y=E$ for ${\cal D}={(\gamma_1,\gamma_2),(d_1,d_2), q}$. The windows $W^{\cal D}_Y$ are given in Table~\ref{tab:filters}. The power spectrum $C^{BB}_\ell({\text cleaned})$ of cleaned B-modes $B_{{\text cleaned}}(\bfl)=B(\bfl)-\hat B(\bfl)$ is thus given by
\begin{eqnarray}
C^{BB}_{\ell}({\text cleaned})= \int  \frac{d^2\bfl_1}{(2\pi)^2} \Big(W^{\cal D}_{Y}(\bfl_1,\bfl_2)\Big)^2  \Bigg[C^{XX}_{\ell_1} C^{\cal D \cal D}_{\ell_2}-\frac{\Big(C^{XX}_{\ell_1}C^{\cal D \cal D}_{\ell_2}\Big)^2 }{\Big(C^{XX}_{\ell_2}+{\cal N}^{XX}_{\ell_2}\Big) \Big(C^{\cal D \cal D}_{\ell_2}+N^{\cal D \cal D}_{\ell_2}\Big)}\Bigg] \,.
\label{eqn:cbbclean}
\end{eqnarray}
We can repeat the above process of cleaning  by using the cleaned $C^{BB}_{\ell}({\text cleaned})$ given by Eqn.~(\ref{eqn:cbbclean}) as the new observed $C^{BB}_{\ell}$ and repeating the cleaning process until $C^{BB}_\ell(cleaned)$ has converged. In each iteration of cleaning process the Gaussian noise of the estimator $N^{\cal D \cal D}_\ell$ reduces until it converges to maximum likelihood estimator noise. The decrease in the estimator Gaussian noise can be understood by noting that the B-mode power spectrum acts as noise source for the quadratic estimator. By iteratively cleaning the B-modes, we are reducing the noise source for the estimator until B-modes are nothing but primordial (i.e. tensor) B-modes.

The uncertainty is the tensor-to-scalar $\Delta r$ is given by
\begin{eqnarray}
\Delta r_{\text{no cleaning}}=\Bigg[\frac{f_{sky}}{2}\sum_\ell (2\ell+1) \Bigg( \frac{C^{BB}_\ell(\text{tensor})}{C^{BB}_\ell(\text{observed})+{\cal N}^{B}_\ell}\Bigg)^2\Bigg]^{-1/2}\,, \\
\Delta r_{\text{with cleaning}}=\Bigg[\frac{f_{sky}}{2}\sum_\ell (2\ell+1) \Bigg( \frac{C^{BB}_\ell(\text{tensor})}{C^{BB}_\ell(\text{cleaned})+{\cal N}^{B}_\ell}\Bigg)^2\Bigg]^{-1/2}\,,
\end{eqnarray}
where $\Delta r_{\text{no cleaning}}$ is the uncertainty in $r$ when no cleaning was performed, while $\Delta r_{\text{with cleaning}}$ is the uncertainty after the distortions have been cleaned. Note that the ratio of uncertainties with and without cleaning is independent of tensor perturbations, and sky fraction covered.

\end{document}